\documentclass{aa}  
\usepackage{lineno}
\renewcommand{\linenumbers}{}
\usepackage{graphicx}
\usepackage{txfonts}
\usepackage[colorlinks=true,allcolors=blue]{hyperref}
\usepackage{lipsum}
\usepackage{subcaption}        
\usepackage{lscape}
\usepackage{placeins}           

\begin{document}

   \title{A joint JWST and HST view of Omega Centauri: Multiple stellar populations and their kinematics}

\author{T.\,Ziliotto$^{1}$\thanks{E-mail: tuila.ziliotto@unipd.it},
       A.\,P.\,Milone$^{1,2}$, G.\,Cordoni$^{3}$, A.\,F.\,Marino$^{2}$, M.\,V.\,Legnardi$^{1}$, E.\,Dondoglio$^{2}$, E.\,Bortolan$^{1}$, F.\,Muratore$^{1}$.
        }

\institute{$^1$  Dipartimento di Fisica e Astronomia ``Galileo Galilei'', Univ. di Padova, Vicolo dell'Osservatorio 3, Padova, 35122, Italy \\
$^{2}$ Istituto Nazionale di Astrofisica - Osservatorio Astronomico di Padova, Vicolo dell'Osservatorio 5, Padova, 35122, Italy\\
$^{3}$ Research School of Astronomy \& Astrophysics, Australian National University, Canberra, ACT 2611, Australia \\
}

   \date{\today}

   \titlerunning{Multiple populations in $\omega$\,Centauri with \textit{HST} and \textit{JWST}.} 
\authorrunning{Ziliotto et al.}

\abstract{We combine F115W and F277W images collected with the Near Infrared Camera of the James Webb Space Telescope with multiband, multi-epoch Hubble Space Telescope observations of $\omega$\,Centauri to investigate its multiple stellar populations and internal kinematics. Our study focuses on a region spanning $\sim$0.9 to $\sim$2.3 half-light radii from the cluster center, which is largely unexplored by these telescopes.
Using chromosome maps, we identified the principal populations along the upper main sequence and among M dwarfs, distinguishing lower-stream stars chemically akin to first-generation globular cluster stars with similar metallicities from upper-stream stars enriched in helium and nitrogen but that are oxygen poor. Both streams also host subpopulations with varying metallicities.
We found radially anisotropic motions, with upper-stream stars exhibiting significantly stronger anisotropy than lower-stream stars. Subdividing the upper stream into extreme and intermediate light-element populations revealed a gradient in anisotropy, with intermediate stars lying between the lower-stream stars and extreme upper-stream populations. However, metal-rich and metal-poor stars within each stream show moderate kinematic differences.
The lower-stream stars show a higher angular momentum dispersion compared to upper-stream stars, and they also exhibit stronger systemic rotation and proper motion skewness, further highlighting their kinematic divergence.
Finally, leveraging a mass range of $\sim$0.15–0.7 M$_\odot$, we detected a low degree of energy equipartition for all cluster stars, which decreases with radial distance from the cluster center.
}


   \maketitle

\section {Introduction}
\label{sec:intro}

As the most massive and enigmatic globular cluster (GC) in the Milky Way, $\omega$\,Centauri is one of the most extensively studied stellar systems in the context of multiple stellar populations. In particular, star-to-star variations in metallicity and light-element abundances have been known for decades \citep[e.g.\,][]{ norris1975a, butler1978a, persson1980a}. $\omega$\,Centauri was also the first GC in which multiple populations were photometrically detected along the red giant branch \citep[RGB; e.g.,][]{wooley1966a, evans1977a}, main sequence \citep[MS;][]{anderson1997a, bedin2004a}, and the sub-giant branch \citep{lee1999a}.

The multiple stellar populations in $\omega$\,Centauri display distinctive chemical abundance patterns as well as features commonly observed in most GCs. Its stars span a wide metallicity range, from [Fe/H] $\lesssim -2$ dex to [Fe/H] $\gtrsim -0.7$ dex \citep{norris1995a, johnson2008a, marino2011a, nitshai2023a}, with metallicity correlating with the abundance of $s$-process elements and the total C$+$N$+$O content \citep{norris1995a, johnson2010a, marino2012a, marino2019a}. Although such star-to-star variations in heavy elements are rare among GCs, stars in $\omega$\,Centauri with similar metallicities show classic GC signatures, including light-element abundance variations such as the C–N, Na–O, and Mg–Al anticorrelations, along with helium enrichment. Among these populations, stars with intermediate metallicities exhibit the most extreme chemical variations across most elements \citep{marino2011a, marino2012a, johnson2010a, clontz2025a}.

The bizarre chemical composition observed in $\omega$\,Centauri is not unique to this GC, as such chemical composition has also been detected in other massive Galactic GCs, including M\,22, M\,2, and NGC\,5286, among others, which exhibit variations in metallicity, $s$-process element abundances, and total C$+$N$+$O content \citep[e.g.,][]{marino2009a, marino2015a, marino2019a, dacosta2009a, yong2014a, mckenzie2022a, dondoglio2025a}. This group of GCs with intrinsic metallicity spreads, known as Type\,II GCs, also includes M\,54, which is located in the nucleus of the Sagittarius dwarf galaxy \citep{carretta2010a}.

Due to its large mass, broad metallicity spread, and retrograde orbit, $\omega$\,Centauri is often considered not to be a genuine GC but rather the remnant nuclear star cluster of a dwarf galaxy that was accreted by the Milky Way or possibly an ultra-compact dwarf galaxy. These scenarios are supported by the fact that $\omega$\,Centauri shares structural properties — such as half-mass radius and total mass — with some known nuclear star clusters and compact dwarf galaxies \citep[e.g.,][]{marino2015a}. Furthermore, recent evidence based on {\it HST} proper motion data has revealed fast-moving stars near the cluster center, supporting the presence of an intermediate-mass black hole \citep{haberle2024a}, as black holes are often found in galaxy nuclei. For these reasons, the investigation of multiple populations in $\omega$\,Centauri may provide valuable insights into not only its star formation history but also into the origin and evolution of Type\,II GCs and, potentially, nuclear star clusters.

Studies based on multiband photometry have highlighted the complexity of multiple stellar populations in $\omega$\,Centauri \citep[e.g.,][]{bellini2010a, bellini2017a}. The distribution of giant stars in the pseudo two-color diagram, known as the chromosome map (ChM), is significantly more complex than what is observed in other GCs, revealing at least 15 distinct groups of stars with different chemical compositions \citep{milone2017a, milone2020a, bellini2017a, lagioia2021a, nitshai2024a}. 

Similar to most Galactic GCs, the $\Delta_{C \rm F275W,F336W,F438W}$ versus \,$\Delta_{F275W,F814W}$ ChM of metal-poor RGB stars in $\omega$\,Centauri displays a group of first-population (1P) stars near the origin of the reference frame along with a sequence of second-population (2P) stars extending toward higher values of $\Delta_{C \rm F275W,F336W,F438W}$. In addition, both 1P and 2P stars form lower and upper streams (hereafter LS and US) of more metal-rich stars, with similar $\Delta_{C \rm F275W,F336W,F438W}$ values that extend to large $\Delta_{F275W,F814W}$ values. Similar to 1P stars, the lower-stream stars are characterized by low abundances of He, N, Al, and Na and high abundances of C and O, in contrast to 2P and upper-stream stars of the same metallicity \citep{milone2017a, marino2019a, clontz2025a}. We note that throughout this work, we use "streams" to refer to these photometric sequences in color-color space, and they are unrelated to spatially coherent stellar streams in the Galactic halo.

The internal kinematics of stars in dynamically young GCs such as $\omega$\,Centauri can retain information about their initial configuration, thus providing valuable constraints on the formation of multiple stellar populations. By combining photometry and proper motions from Gaia Data Release 2 \citep{gaia2018a}, {\it HST}, and ground-based facilities, \citet{cordoni2020a} investigated the kinematics of multiple stellar populations along the RGB of $\omega$\,Centauri, extending out to $\sim$5.5 half-light radii from the cluster center. They found that beyond the half-light radius, US stars exhibit more radially anisotropic motions compared to the LS. These results are consistent with those reported by \citet{bellini2018a}, who used deep {\it HST} photometry and proper motions for stars located at about 3.5 half-light radii from the center of the cluster to show that 2P stars are more radially anisotropic than the 1P. Intriguingly, the anisotropy differences appear less pronounced when comparing stars of different metallicities \citep{cordoni2020a}. A similar conclusion — that stars with different iron abundances share comparable kinematics — is supported by \citet{haberle2025a} based on {\it HST} data within the half-light radius.

In this paper, we use multi-epoch and multiband images collected with the {\it James Webb Space Telescope} ({\it JWST}) and {\it HST} to study multiple stellar populations along the MS and their internal kinematics. Specifically, in Sect.\,\ref{sec:data} we describe the dataset and the methods to derive stellar photometry, astrometry, and proper motions. The photometric diagrams used to detect the multiple stellar populations are presented in Sect.\,\ref{sec:cmds}, whereas their internal kinematics are investigated in Sect.\,\ref{sec:kinematics}. Finally, a summary of the results and a discussion are provided in Sect.\,\ref{sec:summary}.

\section{Data and data reduction}\label{sec:data}
\begin{figure}
    \centering
    \includegraphics[width=0.95\linewidth]{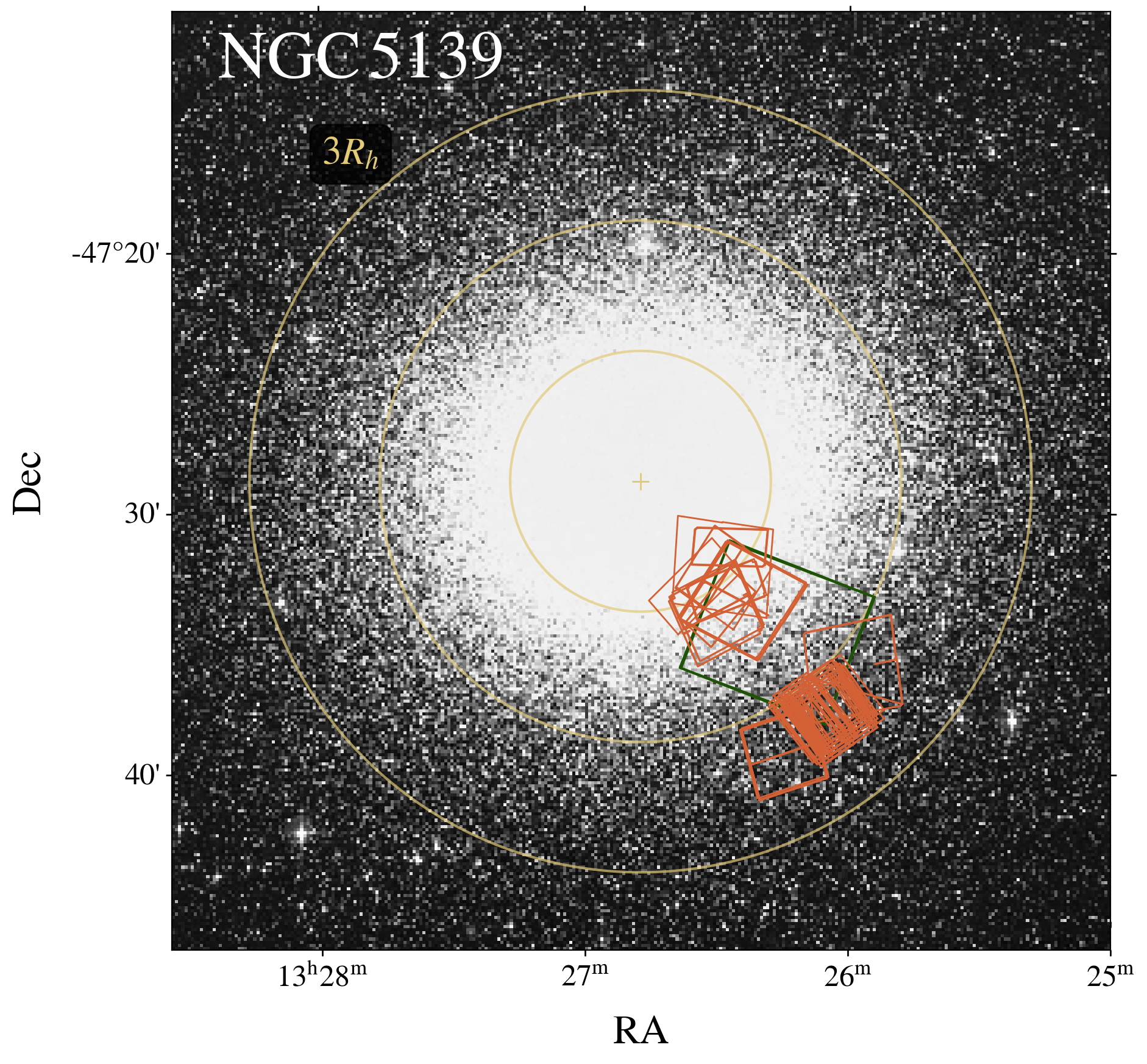}
    \caption{Footprints of the JWST (green) and HST (orange) observations used in this work. North is at the top, and east is to the left. Yellow circles indicate multiples of the half-light radius. The yellow cross indicates the cluster center at coordinates ($\alpha$, $\delta$) = (201.697°, $-47.480^\circ$).}
    \label{fig:footprint}
\end{figure}

Our primary dataset for studying multiple populations in $\omega$\,Centauri comprises NIRCam images taken through the F115W and F277W filters, targeting a field located between $\sim$1 and $2.5$ half-light radii from the cluster center
(GO-6154; PI Anna de Graaff).
This region, situated between the fields analyzed by \citet{haberle2025a} and \citet{bellini2018a}, remains largely unexplored in terms of stellar kinematics. To enhance the identification of multiple stellar populations along the MS and to derive proper motions, we also included archival {\it HST} images from the WFC/ACS, IR/WFC3, and UVIS/WFC3 instruments. These data were obtained from multiple programs (PI Bedin, GO-14118; PI Bellini, GO-15857; PI Renzini, GO-12580; PI Seth, GO-16777; PI Scalco, GO-16247; PI Brown, GO-14759; PI Cool, GO-9442).
The footprint of the archival data used in this work, retrieved from the Mikulski Archive for Space Telescopes (MAST)\footnote{The data is available at \href{http://dx.doi.org/10.17909/6cjp-2k84}{dx.doi.org/10.17909/6cjp-2k84}.}, is shown in Figure~\ref{fig:footprint}. 

\subsection{Astrometry and photometry}
We derived stellar positions and fluxes by applying a two-step photometric and astrometric approach that integrates methodologies optimized for {\it HST} and {\it JWST} imaging. Our analysis makes use of custom-developed software by Jay Anderson, including the {\it img2xym} and KS2 packages \citep[e.g.,][]{anderson2000a, anderson2008a, anderson2022a}.

First-step measurements of stellar fluxes and positions were obtained independently for each image using the {\it img2xym} software, which fits stars using spatially variable point-spread-function (PSF) models. To account for subtle, time-dependent PSF variations, we adopted perturbation PSFs constructed from bright, unsaturated, and isolated stars \citep{anderson2006a}. This allowed for more accurate modeling of each exposure, particularly in the presence of focus variations or temporal changes in the instrument’s optics.

The magnitudes from individual exposures were brought onto a common photometric scale by referencing them to the deepest exposure in each filter. This was achieved by matching well-measured stars across exposures and computing the average magnitude offsets, which were then applied to align all photometry to a common zero point. This reference system served as the master photometric frame.

Refined photometric and astrometric measurements were carried out using the KS2 software package, an advanced extension of the original {\it kitchen\_sync} algorithm \citep{anderson2008a}. KS2 employs three distinct photometric techniques tailored to stars of different brightness and crowding levels. Method I is optimized for bright stars with prominent peaks. After subtracting neighboring sources, stars are measured within a 5$\times$5-pixel region using an effective PSF tailored to their location on the detector. The background sky is estimated from an annulus between 4 and 8 pixels from the stellar center. Method II is designed for fainter stars that may not be individually detectable in every image. It combines data from all exposures to perform weighted aperture photometry within a 5$\times$5 grid, down-weighting pixels contaminated by neighbors. Sky estimation in Method II is performed identically to Method I. Method III is used for the faintest stars in highly crowded fields. It performs aperture photometry using a 0.75-pixel-radius aperture, with the sky background measured in a tighter annulus between 2 and 4 pixels.

Each star's final position and flux were computed by averaging independent measurements across all available exposures. Quality parameters generated by KS2 (e.g., sharpness, crowding, and fit quality) were used to select stars with reliable PSF fits following the procedures described in \citet[][Section 2.4]{milone2023a}.
Geometric distortion corrections were applied to all images using the most recent solutions for {\it HST} and {\it JWST} instruments \citep{anderson2006a, bellini2009a, bellini2011a, anderson2022a, milone2023b}. All stellar positions were registered onto a common astrometric frame aligned with Gaia data release 3 \citep[DR3,][]{gaia2021a}, ensuring consistency with celestial coordinates (North up, East left).
Photometric calibration was performed in the Vega magnitude system using the latest zero points from STScI for {\it HST} and {\it JWST} filters\footnote{\href{https://www.stsci.edu/hst/instrumentation/wfc3/data-analysis/photometric-calibration}{https://www.stsci.edu/hst/instrumentation/wfc3/data-analysis/photometric-calibration};\\
\href{https://jwst-docs.stsci.edu/jwst-near-infrared-camera/nircam-performance/nircam-absolute-flux-calibration-and-zeropoints}{https://jwst-docs.stsci.edu/jwst-near-infrared-camera/nircam-performance/nircam-absolute-flux-calibration-and-zeropoints}} \citep{milone2023b}. Extensive artificial-star experiments, executed with the KS2 framework, were used to assess completeness, crowding effects, and photometric uncertainties \citep{anderson2008a, milone2023a}.
We corrected the photometry for the effects of differential reddening by using the procedure and the computer programs widely used in previous work from our group
\citep{milone2012a, jang2022a, legnardi2023a}.

\subsection{Proper motions}

To measure relative proper motions, we adopted a method commonly used in prior studies \citep[e.g.,][]{anderson2003a, piotto2012a, libralato2022a, milone2023a, ziliotto2025a} based on comparing stellar positions across multi-epoch images.
We began with independent photometric and astrometric catalogs from various filters and epochs. Our reference frame was defined by NIRCam F115W images, oriented using Gaia DR3 data, with the X-axis pointing west and the Y-axis pointing north.

Stellar positions from each catalog were transformed into this master frame using six-parameter linear fits. To correct for residual distortions, we applied local transformations based on the 100 nearest reference stars with similar luminosities, excluding the target star itself.
Proper motions were derived by fitting a weighted least-squares line to each star’s position as a function of time. The slope and its uncertainty represent the proper motion and its error.

Bright, unsaturated cluster stars, selected following the criteria in the previous subsection and \cite{milone2023a}, were used for transformations. An initial selection was based on CMD position, followed by refinement using both CMD and proper motion diagrams. These members were used to compute final motions.

\begin{figure}
    \centering
    \includegraphics[width=1\linewidth]{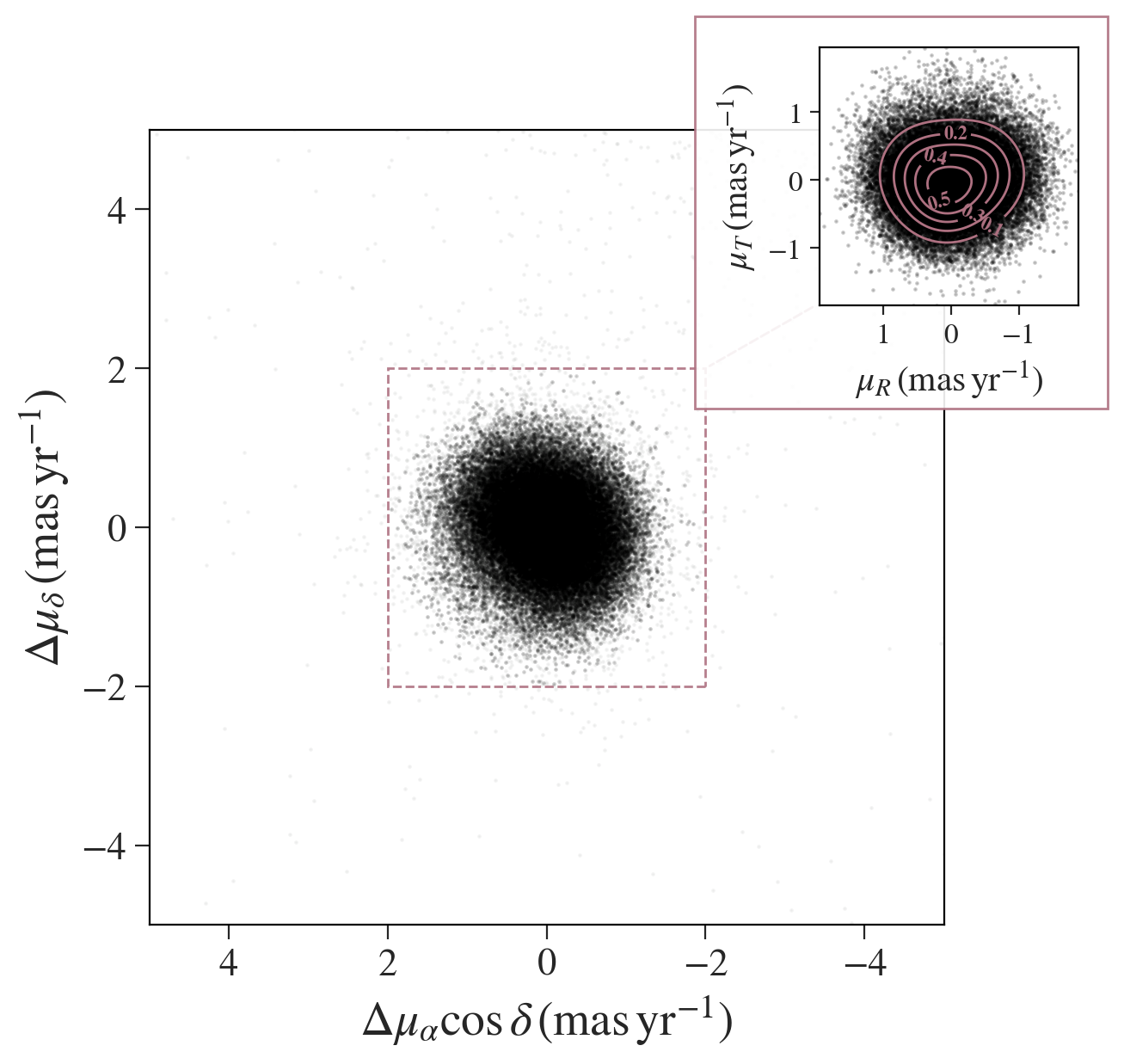}
    \caption{Distribution of proper motions measured in the cluster reference frame. The inset panel displays the projected proper motions in tangential and radial directions. Pink contours indicate probability density distributions derived from Gaussian mixture model fitting.}
    \label{fig:pms_all}
\end{figure}

To enhance accuracy, we corrected each star’s position at every epoch for its proper motion, recalculated the transformations, and derived improved proper motions.
Finally, we converted positions and proper motion components from equatorial to Cartesian coordinates following Equation 2 from \cite{gaia2018b}. Then, we projected these proper motions into radial ($\mu_\mathrm{R}$) and tangential ($\mu_\mathrm{T}$) components. 
 The resulting proper motions relative to the mean motion of $\omega$\,Centauri in the $\mu_{\delta}$ versus\,$\mu_{\alpha} {\rm cos} \delta$  plane are illustrated in Figure\,\ref{fig:pms_all}, whereas the inset shows the same proper motions in the tangential versus radial direction with respect to the cluster center located at ($\alpha$, $\delta$) = (201.697°, $-47.480^\circ$), as in \citet{vasiliev2021}.

 Following photometric and astrometric quality cuts, our sample contains 29,676 stars in the innermost intersection of JWST and HST pointings and 10,686 stars in the outermost intersection, with median proper motion uncertainties of 0.04 and 0.06 $\rm mas \,yr^{-1}$, respectively. The corresponding observation time baselines span up to $\rm \sim 23 \, yr$ and $\rm \sim 13 \, yr$, respectively.

\section{Multiple populations along the main sequence}\label{sec:cmds}
\begin{figure*}
    \centering
    \includegraphics[width=0.85\linewidth]{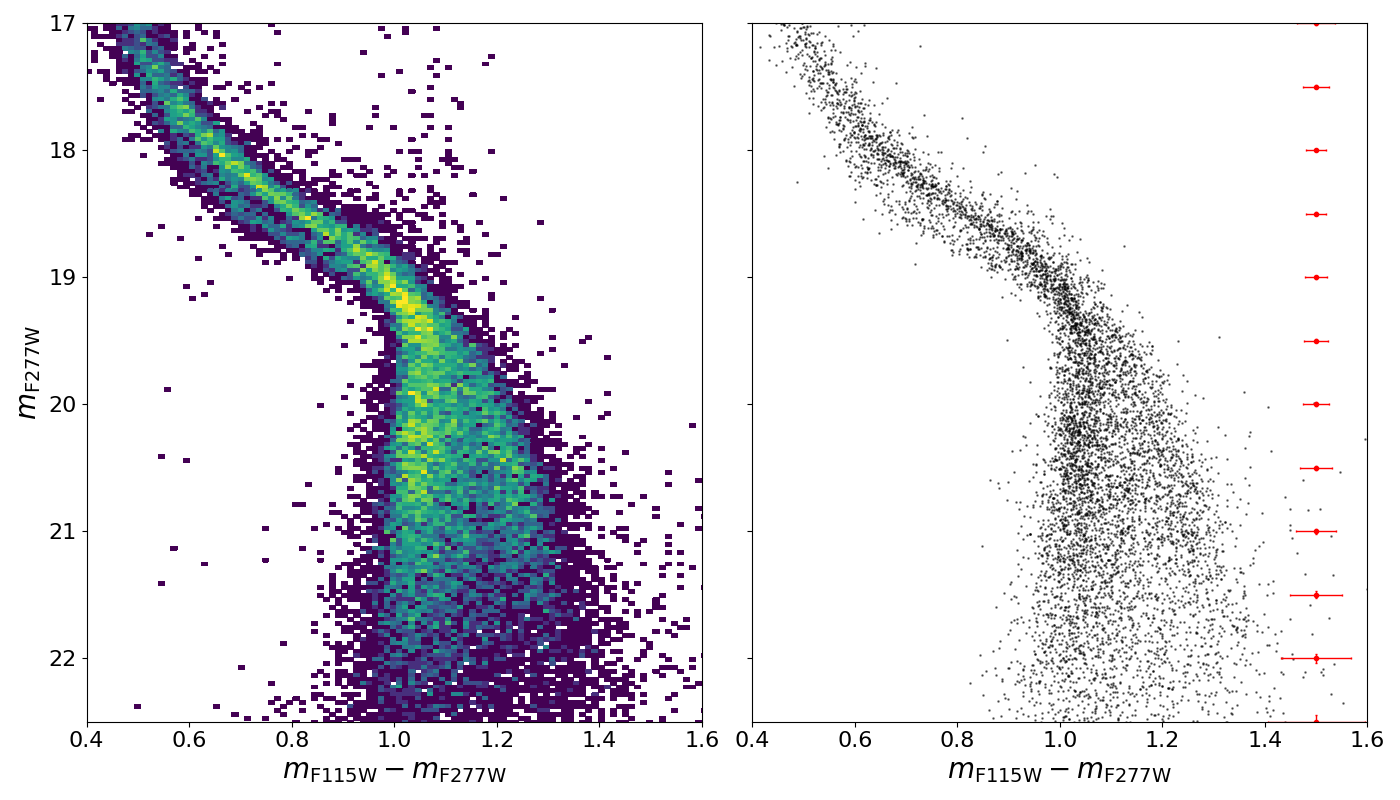}
    \caption{$m_{\rm F277W}$ vs.\,$m_{\rm F115W}-m_{\rm F277W}$ Hess diagram for stars at $r_h > 1.5$ that according to the criteria of Section\,\ref{sec:data} have high-quality photometry (left). The right panel shows the corresponding CMD for proper-motion selected cluster members.}
    \label{fig:CMD115277}
\end{figure*}

Figure \ref{fig:CMD115277} shows the Hess diagram of all stars with high-quality photometry, selected according to the criteria described in Section \ref{sec:data}, in the $m_{\rm F277W}$ versus \ $m_{\rm F115W} - m_{\rm F277W}$ plane. The right panel presents the corresponding CMD for proper-motion-selected cluster members located at radial distances larger than approximately 1.5 half-light radii from the cluster center. 

In the context of multiple stellar populations, this CMD resembles the $m_{\rm F160W}$ versus \ $m_{\rm F110W} - m_{\rm F160W}$ CMD based on IR/WFC3 photometry \citep{milone2017b}. One of the most striking features in the upper part of the CMD is the split MS, first identified by \citet{anderson1997a}. The blue MS consists of stars with extreme helium enhancement (Y$\gtrsim$0.38), while the red MS is populated by stars with more canonical helium abundances \citep[Y$\sim$0.25; e.g.,][]{norris2004a, king2012a, tailo2015a}. These two sequences merge near the MS knee but diverge again at fainter magnitudes. Notably, below $m_{\rm F277W} \sim 19.5$ mag, most red-MS stars form the bluest sequence in the CMD \citep[see Section 4.2 of][for a discussion of these features in an analogous CMD]{milone2017b}.

The MS splitting below the knee primarily reflects differences in oxygen abundance among the populations. The spectra of oxygen-rich M dwarfs exhibit strong absorption beyond $\sim$13,000\AA, primarily due to oxygen-bearing molecules such as water vapor. These molecular features significantly affect the flux in the F277W band, while their impact on the F115W band is less pronounced. As a result, oxygen-poor M dwarfs appear bluer in $m_{\rm F115W} - m_{\rm F277W}$ than oxygen-rich stars of similar luminosity \citep{milone2012b, milone2017b, milone2019a, gerasimov2022a, ziliotto2023a, cadelano2023a, marino2024a, marino2024b}.

\begin{figure*}
    \centering
    \includegraphics[width=0.45\linewidth]{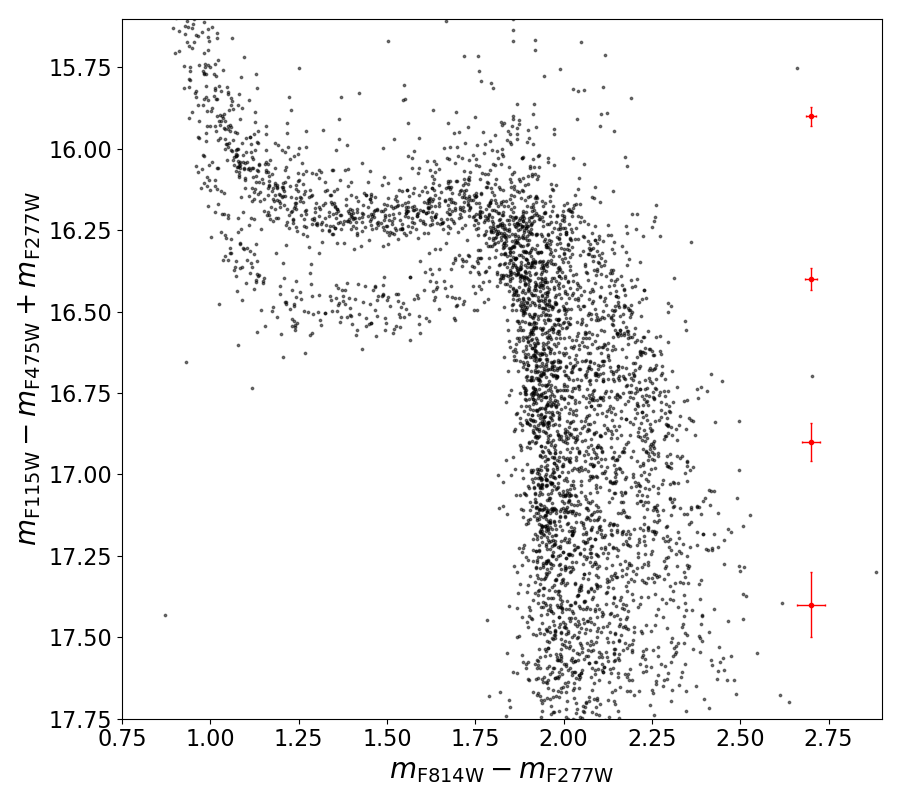}
    \includegraphics[width=0.45\linewidth]{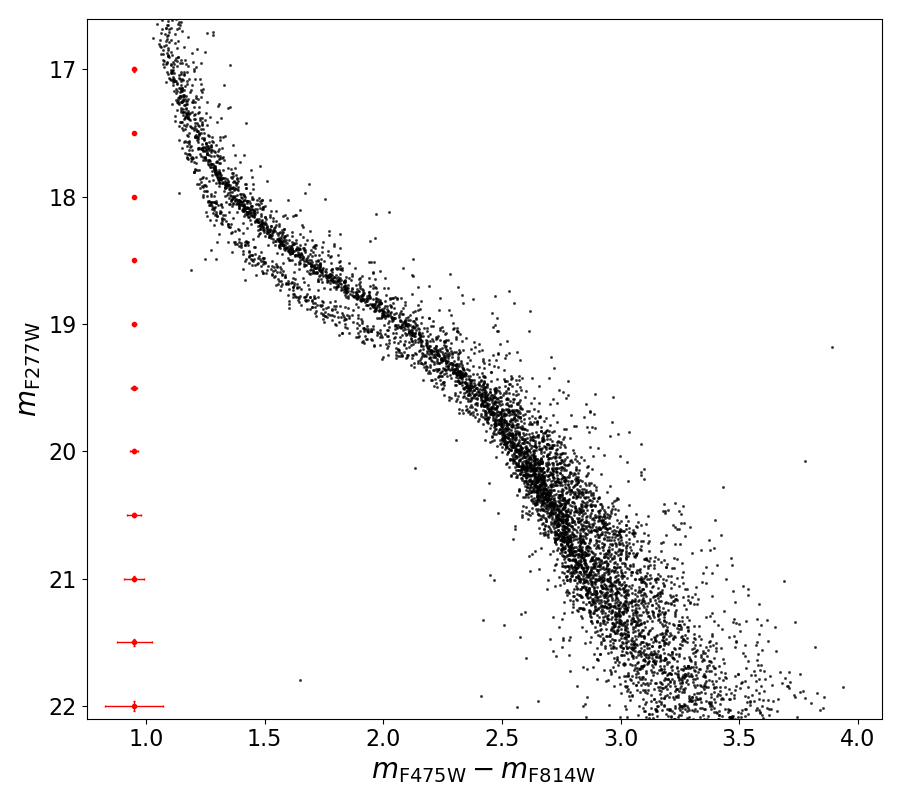}
    \includegraphics[width=0.45\linewidth]{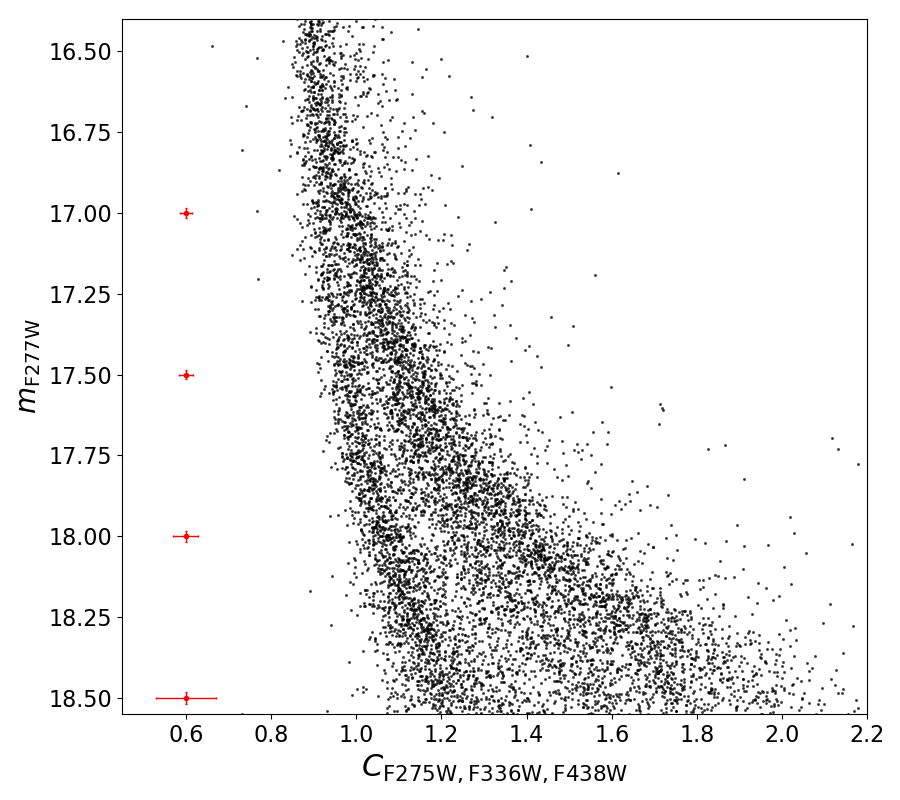}
    \includegraphics[width=0.45\linewidth]{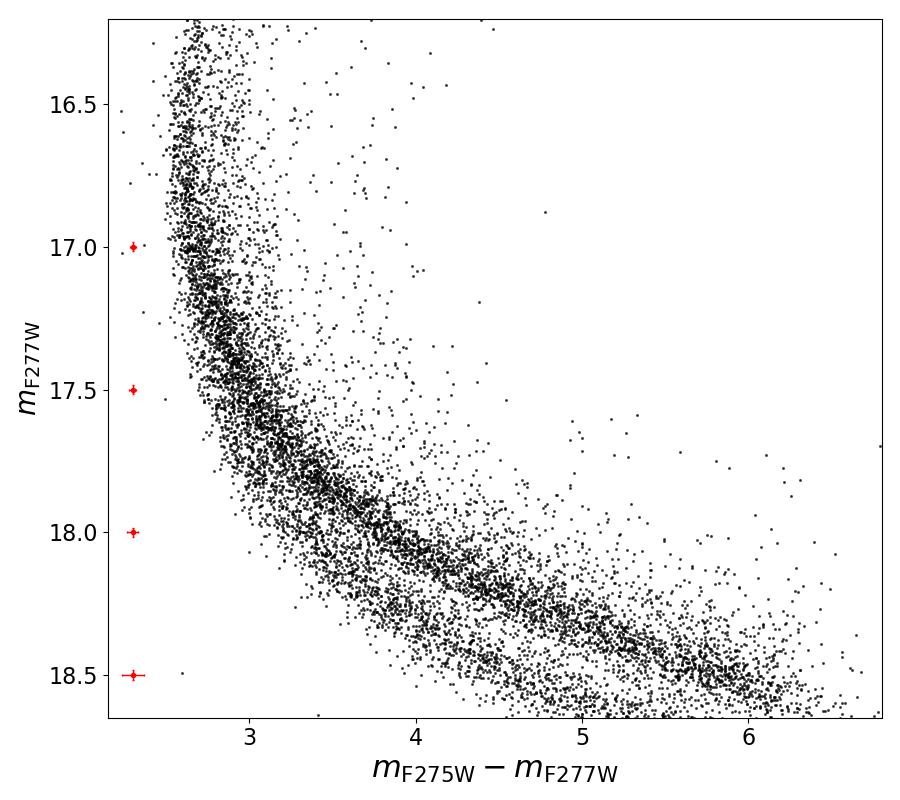}
    \caption{Collection of photometric diagrams corrected for differential reddening that highlight the multiple MSs of $\omega$\,Centauri proper-motion selected members.}
    \label{fig:CMDs}
\end{figure*}

The synergy of photometry from UV, optical, and NIR bands can provide further details on the multiple MSs of $\omega$\,Centauri.
Figure\,\ref{fig:CMDs} presents four CMDs based on both HST and JWST photometry, each revealing clear evidence of multiple stellar populations along the MS. The top-left panel shows the pseudo-CMD constructed from $m_{\rm F115W}-m_{\rm F475W}+m_{\rm F277W}$ versus $m_{\rm F814W}-m_{\rm F277W}$, which highlights the separation between the red and blue MS components above the MS knee. This behavior is analogous to that observed in CMDs built from similar pseudo-magnitudes \citep[e.g.,][]{milone2019a}. Notably, the diagram exhibits an abrupt change in the slope of the MS near the knee, making it a particularly effective diagnostic for accurately identifying this key evolutionary feature.

The wide color baseline of the $m_{\rm F277W}$ versus \,$m_{\rm F475W}-m_{\rm F814W}$ CMD, shown in the upper-right panel of Figure \ref{fig:CMDs}, serves as an effective diagnostic for distinguishing multiple stellar populations with varying helium abundances \citep[e.g.\,][]{dantona2005a, piotto2007a, dantona2022a}. While the red and blue MSs appear as distinct stellar populations along the upper MS, their separation becomes less evident below the MS knee — consistent with the trend observed in the top-left CMD. Interestingly, this diagram also reveals a sparsely populated sequence of M dwarfs with redder colors than the main MS locus at similar luminosities. These stars are believed to be highly metal-rich, as previously reported by \citet{milone2017b}.

The $m_{\rm F277W}$ versus \,$m_{\rm F275W}-m_{\rm F277W}$ diagram shown in the bottom-right panel of Figure \ref{fig:CMDs} reveals an even greater separation between stellar populations with different helium abundances. This enhanced separation arises because, at fixed luminosity, helium-rich stars are significantly hotter than their helium-normal counterparts. Additionally, the $m_{\rm F275W}-m_{\rm F277W}$ color is highly sensitive to differences in effective temperature, making it an excellent diagnostic for distinguishing between populations with similar colors but varying temperatures \citep[e.g.,][]{dantona2002a, milone2023a, ziliotto2023a}. 

Finally, we show in the bottom-left panel of Figure \ref{fig:CMDs} the $m_{\rm F277W}$ versus \,$C_{\rm F275W,F336W,F438W}$ pseudo-CMD for bright MS stars, where the $C_{\rm F275W,F336W,F438W}=$($m_{\rm F275W}-m_{\rm F336W}$)$-$($m_{\rm F336W}-m_{\rm F438W}$) pseudo color maximizes the separation among sequences of stars with different nitrogen abundances \citep{milone2013a}.

\begin{figure}
    \centering
    \includegraphics[width=1.0\linewidth]{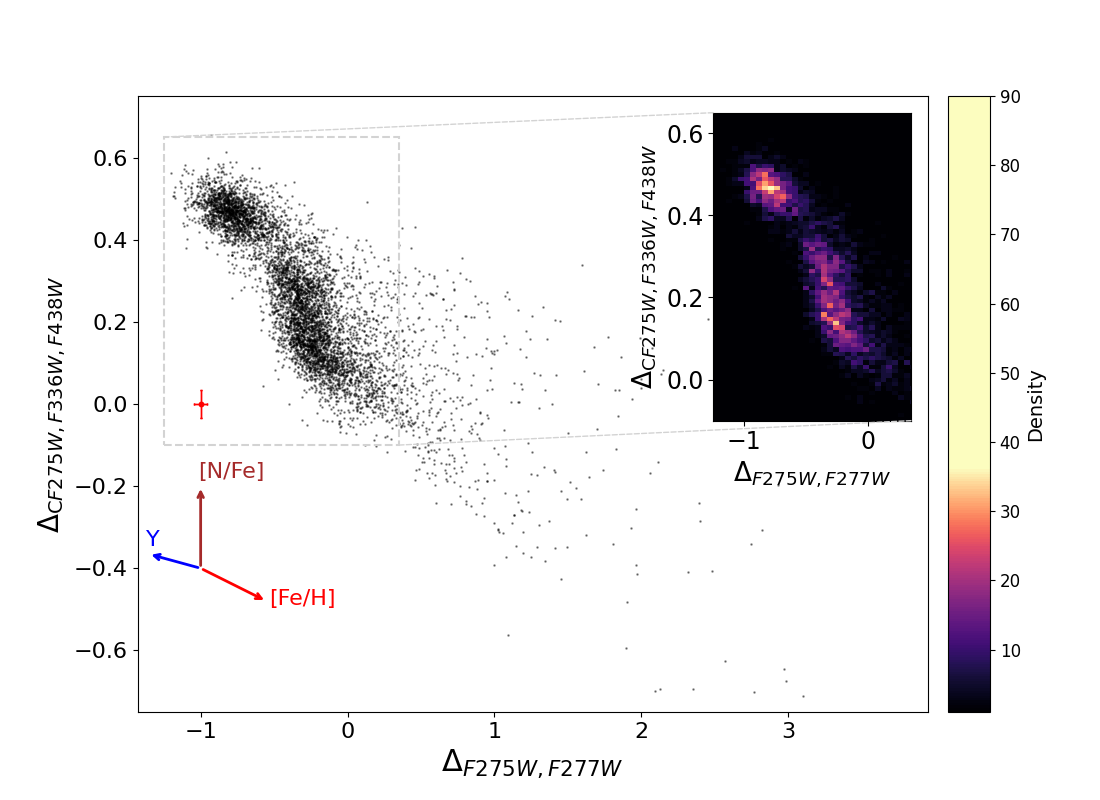}
    \includegraphics[width=1.0\linewidth]{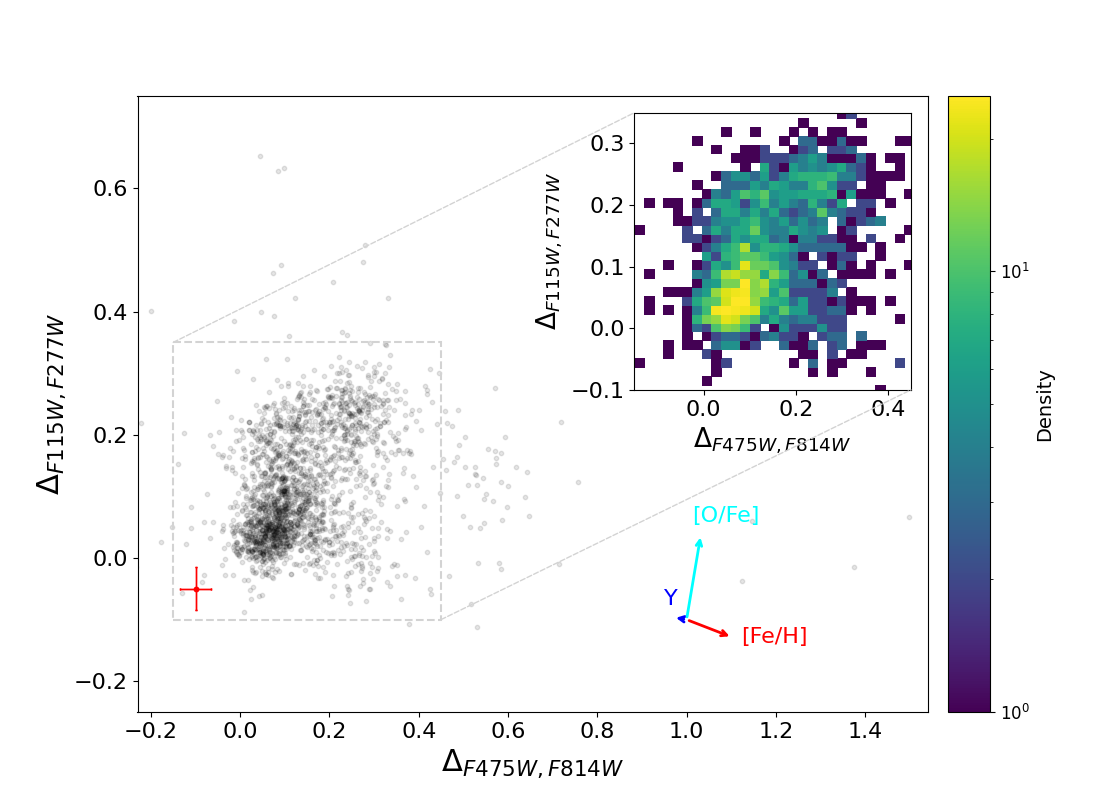}
    \caption{$\Delta_{C {\rm F275W,F336W,F438W}}$ vs.\,$\Delta_{\rm F275W,F277W}$ ChM for bright MS stars (top) and $\Delta_{\rm F115W,F277W}$ vs.\,$\Delta_{\rm F475W,F814W}$ for M dwarfs (bottom). The insets show the corresponding Hess diagrams. The arrows show the effect of changing, one at time,  the abundances of helium mass fraction Y, [N/Fe], [O/Fe], and [Fe/H] by 0.05, 1.2, $-$0.5, and 0.3 dex with respect to a reference stellar population with [Fe/H]=$-$1.7, [$\alpha$/Fe]=0.4 and solar carbon and nitrogen contents \citep[see][for details]{milone2018a, marino2019a}.}
    \label{fig:ChMup}
\end{figure}

To better identify stellar populations along the bright and faint MS segments, we used the CMDs shown in the bottom panels of Figure \ref{fig:CMDs}, which were employed to derive the ChM for the bright MS, and those in Figure \ref{fig:CMD115277} and in the upper-right panel of Figure \ref{fig:CMDs}, used to derive the ChM for the faint MS.
 The ChM is a photometric diagram analogous to a two-color diagram, but it employs pseudo-colors obtained by verticalising the MS relative to its boundaries \citep[see][for details]{milone2017a}.
The results are illustrated in Figure \ref{fig:ChMup}. The upper panel shows the $\Delta_{C\,\mathrm{F275W,F336W,F438W}}$ versus $\Delta_{\mathrm{F275W,F277W}}$ ChM for bright MS stars (17.6 $< m_{\rm F277W} < 18.6$). The stellar distribution is qualitatively similar to that observed in the $\Delta_{C\,\mathrm{F275W,F336W,F438W}}$ versus $\Delta_{\mathrm{F275W,F814W}}$ ChM of RGB stars. Indeed, in both ChMs, the stellar positions are primarily governed by their relative abundances of nitrogen, helium, and metals, as illustrated by the arrows 
that show the effect on pseudo-colors of individually varying the helium mass fraction Y, [N/Fe], [O/Fe], and [Fe/H] by 0.05, 1.2, $-$0.5, and 0.3 dex, respectively, relative to a reference population with [Fe/H] = $-$1.7, [$\alpha$/Fe] = 0.4, and solar carbon and nitrogen abundances
\citep{milone2012b, milone2017a, milone2020a, marino2019a, marino2024a,marino2024b, ziliotto2023a}.
The majority of metal-poor stars, characterized by helium abundances of Y~$\sim$~0.25 and nearly solar nitrogen content, are clustered near the origin of the diagram. In contrast, the bulk of stars with enhanced helium and nitrogen abundances form a distinct sequence extending toward higher values of both $\Delta_{C\,\mathrm{F275W,F336W,F438W}}$ and $\Delta_{\mathrm{F275W,F277W}}$. In this ChM, metal-rich stars delineate stellar streams
that extend toward the lower-right region of the diagram.

The $\Delta_{\mathrm{F115W,F277W}}$ versus $\Delta_{\mathrm{F475W,F814W}}$ ChM shown in the bottom panel of Figure \ref{fig:ChMup} highlights M dwarfs (19.8 $< m_{\rm F277W} <21.7$) with different chemical compositions. Helium-poor, metal-poor stars cluster near the origin of the diagram, while metal-poor stars with extreme light-element abundances are located around ($\Delta_{\mathrm{F115W,F277W}}$, $\Delta_{\mathrm{F475W,F814W}}$)$\sim$(0.3, 0.25). Metal-rich stars are distributed toward the lower-right region of the ChM. 

\begin{figure}
    \centering
    \includegraphics[width=1.0\linewidth]{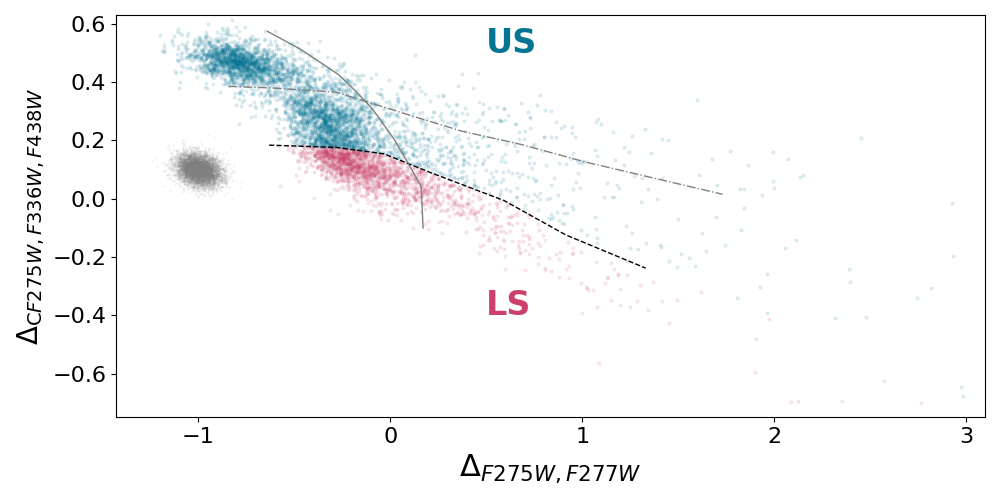}
        \includegraphics[width=1.0\linewidth]{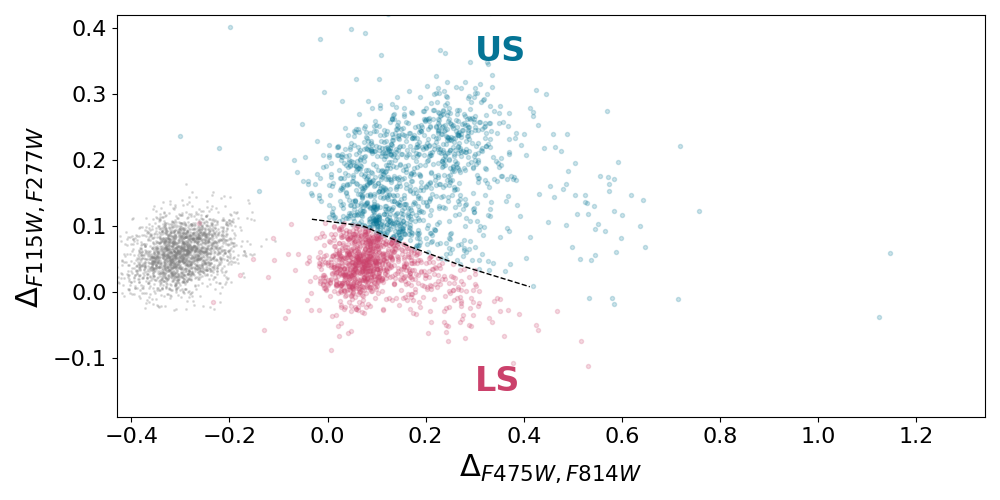}
    \caption{Reproduction of the ChMs from Figure \ref{fig:ChMup}, with bright MS stars (17.6 $< m_{\rm F277W} < 18.6$) in the top panel and M dwarfs (19.8 $< m_{\rm F277W} <21.7$) in the bottom panel. LS and US stars are shown in red and blue, respectively. The black dashed line separates the selected US and LS populations. The dashed-dotted and solid gray lines mark the boundaries between the upper and middle streams, and between metal-rich and metal-poor stars, respectively. The gray dots represent the simulated ChMs for a simple stellar population, and are shifted to the left for clarity.}
    \label{fig:ChMsel}
\end{figure}

We used these ChMs to select the bulk of LS and US stars, as shown in Figure \ref{fig:ChMsel} \citep[e.g.,][]{marino2019a}.
To derive the boundaries between the LS and US populations in the upper panel of Figure \ref{fig:ChMsel}, we adapted to the ChM of $\omega$ Centauri the method introduced by \citet[][see their Figure 2]{milone2017b}. This approach compares the pseudo-color distribution of a simulated simple stellar population (gray points in Figure \ref{fig:ChMsel}) with that of the observed stars. Specifically, we divided the x-axis into bins 0.25 mag wide. For each bin, we shifted the ChM of the simulated population so that the lower tails of the observed and simulated $\Delta_{C\,F275W,F336W,F438W}$ distributions overlap. We then identified the $\Delta_{C\,F275W,F336W,F438W}$ value that encloses 99\% of the simulated points and associated it with the corresponding mean $\Delta_{F275W,F277W}$ value in the ChM. The points obtained for the different bins were linearly interpolated to define the black dashed line. A similar procedure was applied to derive the boundary between the upper and middle streams, and to identify the main loci of LS and US stars in the bottom-panel ChM. Finally, the boundary separating the two main metallicity groups was drawn by eye, to isolate the bulk of metal-poor stars — clearly clustered on the left side of the ChM — from the metal-poor tail. We cross-matched our catalog with the publicly available \cite{nitshai2023a} MUSE catalog, confirming that stars on the blue sequence have an average $\rm [M/H] = -1.50$, while those on the red sequence have an average $\rm [M/H] = -1.24$, consistent with the two sequences tracing distinct metallicity populations. Our final sample comprises 3252 LS stars and 4712 US stars. To evaluate the spatial distribution of these populations, we examined the US-to-LS ratio in two regions: one at $\sim 1 R_h$ and another at $\sim 2 R_h$. The inner region shows a ratio of 1.98, while the outer region has a ratio of 0.75, suggesting that US stars are concentrated toward the cluster center, whereas LS stars become more prevalent in the outer regions.
In addition, we manually defined the dashed gray line, which separates stellar streams with intermediate and extreme light-element abundances (hereafter referred to as the intermediate and extreme streams), and the solid gray line, which distinguishes the majority of metal-rich from metal-poor stars. This classification yields 2282 extreme stream stars and 1576 intermediate stream stars in our final sample.
Based on metallicity classification, our sample comprises 1638 LS metal-poor stars, 460 LS metal-rich stars, 3337 US metal-poor stars, and 522 US metal-rich stars.

\section{Internal kinematics of the multiple populations}\label{sec:kinematics}

\subsection{Velocity dispersion and anisotropy profiles}\label{subsec:ani}

In order to derive the velocity dispersion profiles, we followed the methodology described in \cite{ziliotto2025a}. 
 We first divided the field of view into a series of concentric annular regions, each including approximately the same number of stars. For each annulus we determined the velocity dispersion in the radial and tangential components by maximizing the log-likelihood function described in Equation 1 of \cite{libralato2022a}. We determined the uncertainties in the velocity dispersions using the Markov chain Monte Carlo (MCMC) algorithm \textit{emcee} \citep{foreman2013}.

Moreover, we fit the velocity dispersion and anisotropy profiles following the procedure described in, for example, \cite{aros2025,ziliotto2025a}. Based on the results of simulations of multiple population evolution, we modeled the velocity anisotropy as

\begin{equation}
\beta(R) = 1-\frac{\sigma_\mathrm{T}^2}{\sigma_\mathrm{R}^2} = \frac{\beta_{\infty}R^2}{R_a^2+R^2}\left(1-\frac{R}{R_\mathrm{t}}\right),
\label{eq:ani}
\end{equation}

where $\sigma_\mathrm{T}$ and $\sigma_\mathrm{R}$
are tangential and radial velocity dispersions,
$R_\mathrm{a}$ is the anisotropy radius, $\beta_{\infty}$ is the anisotropy at $R \gg R_\mathrm{a}$, and $R_\mathrm{t}$ is the truncation radius where velocities return to isotropy \citep[see][for more details]{dalessandro2024,aros2025,ziliotto2025a}.

We modeled the radial velocity dispersion using a cubic polynomial, $\sigma_{R}(R) = c_\mathrm{0}+c_\mathrm{1}R+c_\mathrm{2}R^2+c_\mathrm{3}R^3$ \citep{watkins2022}. The coefficients must satisfy two physical constraints: $c_\mathrm{0}+c_\mathrm{1}R+c_\mathrm{2}R^2+c_\mathrm{3}R^3 \geq 0$ and $c_\mathrm{1}+2c_\mathrm{2}R+3c_\mathrm{3}R^2 < 0$. The tangential velocity dispersion follows from the anisotropy relation: $\sigma_\mathrm{T}^2 = (1-\beta(R | R_\mathrm{a},\beta_{\infty},R_\mathrm{t}))\sigma_\mathrm{R}^2$.
We determine the best-fit model parameters through MCMC analysis of the observed individual radial and tangential proper motions.

\begin{figure}
    \centering
    \includegraphics[width=0.88\linewidth]{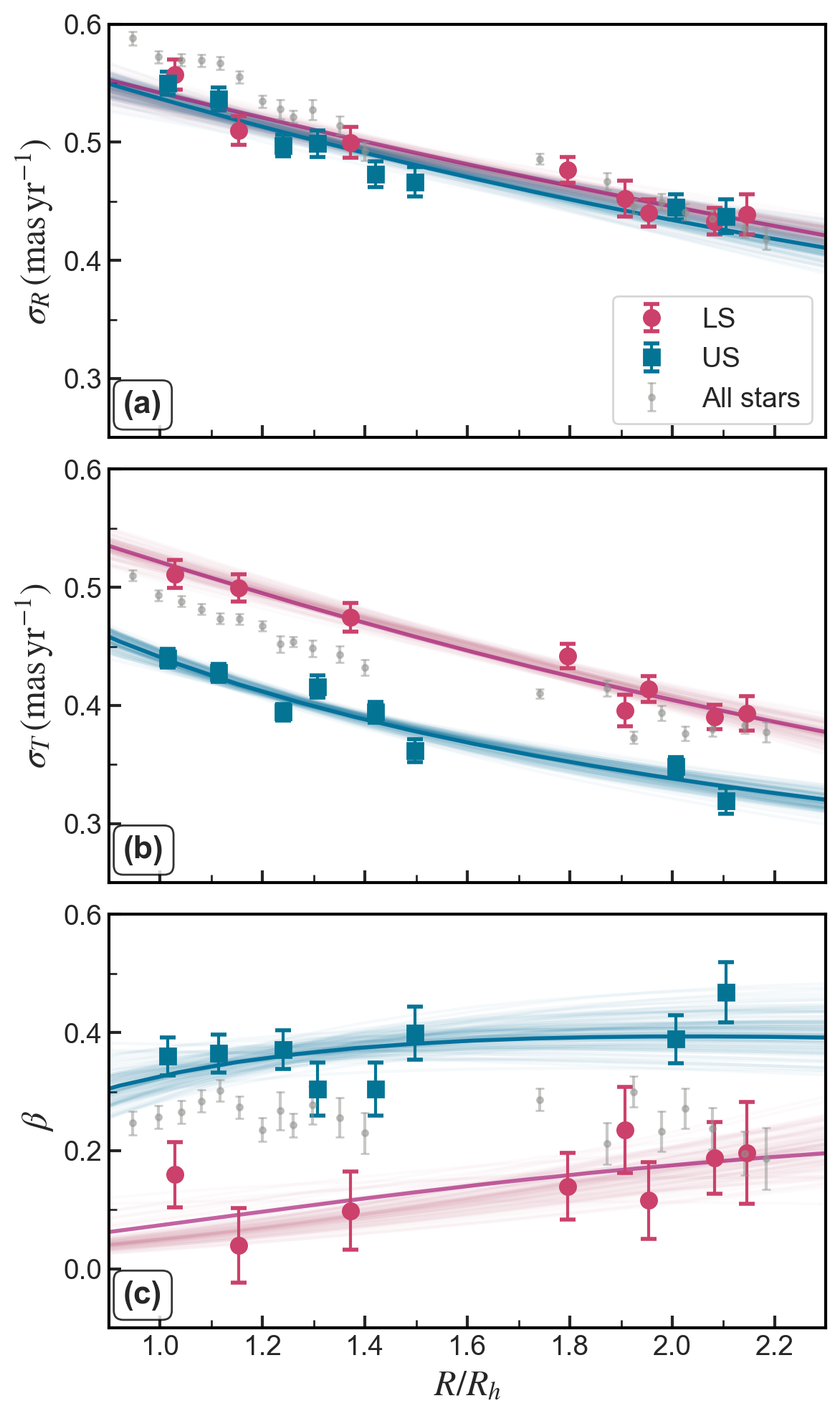}
    \caption{Radial profiles of the velocity dispersion in the radial (a) and tangential (b) components. Panel (c) shows the anisotropy profiles, with $\beta = 1 - (\sigma_T/\sigma_R)^2$. Blue squared markers represent the US population, red dots the LS population, and gray points all stars. The best-fit models for US and LS stars are indicated by dark blue and red lines, respectively. The radial coordinate is normalized to the half-light radius, $R_h = 5.0$ arcmin.}
    \label{fig:ani}
\end{figure}

Figure \ref{fig:ani} presents the velocity dispersion and anisotropy profiles for stars in $\omega$\,Centauri. The top and middle panels show radial and tangential velocity dispersions, respectively, while the bottom panel displays the anisotropy profile. Red circles represent LS stars, blue squares represent US stars, and gray points show the data for all stars. We used bins containing the same number of stars, which are displayed for visual comparison only and not used for fitting. Solid lines indicate the best-fit models for each population (red for LS, blue for US), and non-bolded lines represent additional model realizations based on sampled parameter sets. Radial distances are normalized to the half-light radius (5.0 arcminutes; \citealt[][2010 version]{harris1996a}).

The two populations exhibit similar radial velocity dispersions but differ markedly in their tangential components. LS stars show consistently higher tangential velocity dispersion than US stars across the analyzed radial range ($\sim 1 - 2 R_h$). This tangential difference drives distinct anisotropy behaviors: while both populations are radially anisotropic, US stars display significantly stronger radial anisotropy than LS stars, which remain nearly isotropic in the inner regions.

\begin{figure}
    \centering
    \includegraphics[width=0.88\linewidth]{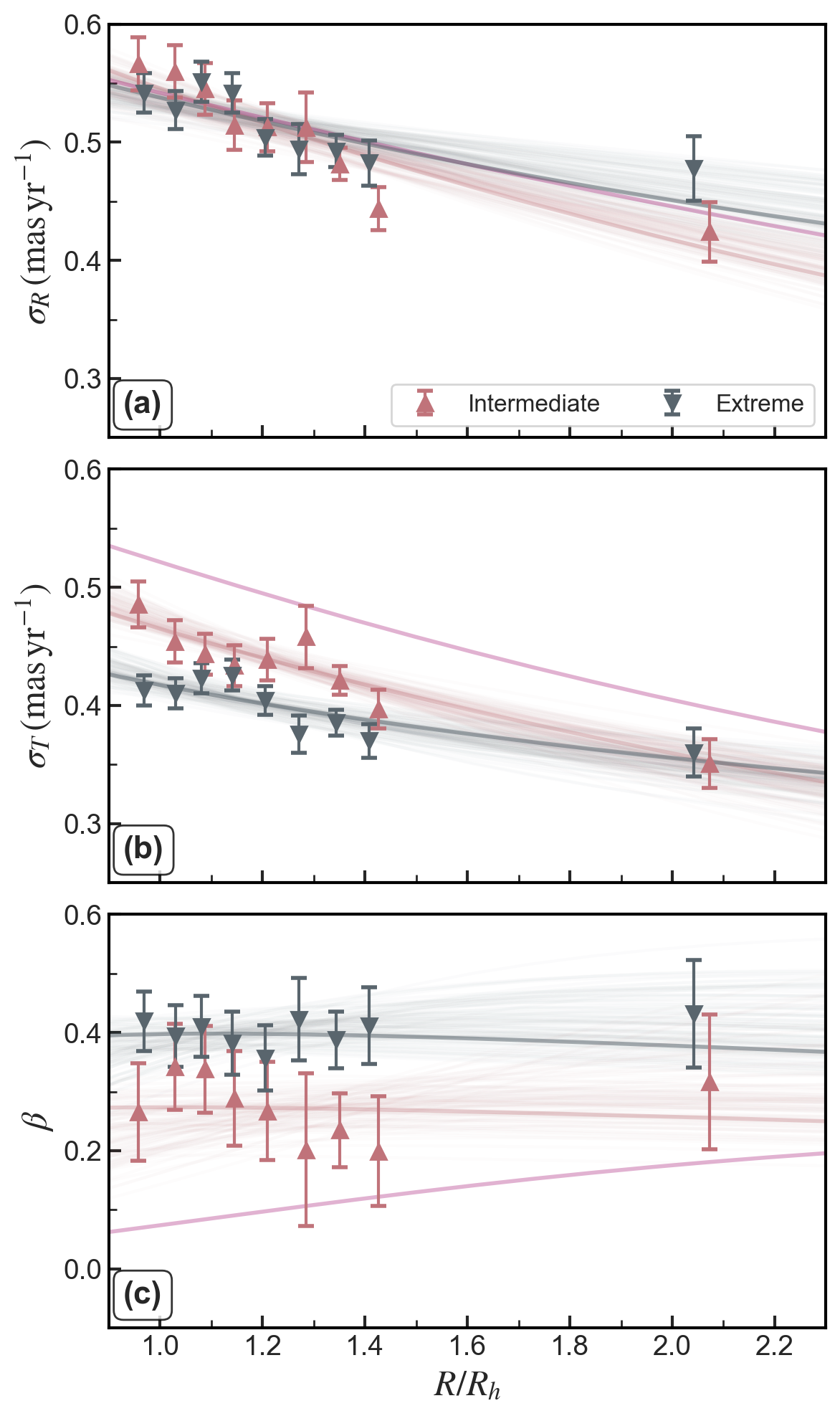}
    \caption{Similar to Figure \ref{fig:ani}, but comparing the chemically intermediate and extreme streams (pink and gray triangles with their corresponding colored lines, respectively). The fit for LS stars (red line) is displayed for comparison.}
    \label{fig:subpops}
\end{figure}

We further analyzed US bright MS stars by dividing them into two subpopulations, a more chemically extreme one, which is located in the uppermost region of the ChM, and an intermediate stream, corresponding to the region located above the LS sequence in the ChM (see Figure \ref{fig:ChMsel}). Similarly to our analysis of US and LS stars, we compare the velocity dispersion and anisotropy profiles of these subpopulations in Figure \ref{fig:subpops}. Pink triangles represent the intermediate stream, while gray triangles represent the extreme stream. The dark solid lines show the best fit for each subpopulation, and the red line shows the best fit for the LS population, which exhibits the most isotropic behavior.
The results show that the extreme stream exhibits the highest radial anisotropy values, while the intermediate stream falls between the two extremes, displaying higher anisotropy than LS stars but lower than the extreme stream. The LS population remains the most isotropic of the three groups.

\begin{figure*}
    \centering
    \includegraphics[width=0.74\linewidth]{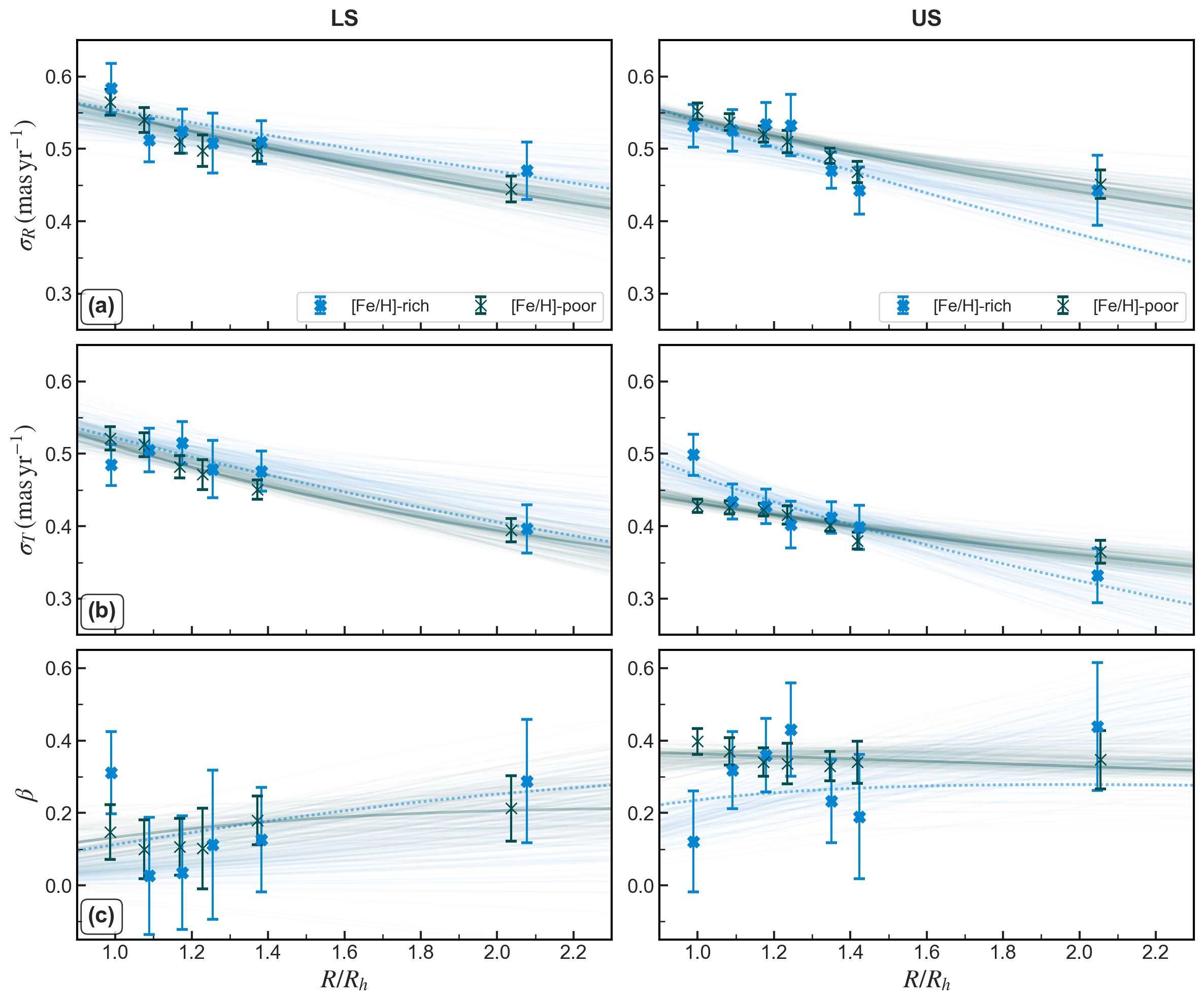}
    \caption{Same as Figure \ref{fig:ani}, but showing separate profiles for metal-rich (blue) and metal-poor (green) stars within each population. Left columns show LS stars, right columns show US stars.}
    \label{fig:metalrichpoorkin}
\end{figure*}

We extended our analysis to investigate metallicity effects by subdividing both LS and US stars into metal-rich and metal-poor subgroups. Figure \ref{fig:metalrichpoorkin} presents the velocity dispersion and anisotropy radial profiles for these metallicity-based subpopulations. The left panels display results for LS stars, showing radial velocity dispersion (top), tangential velocity dispersion (middle), and anisotropy (bottom). The right panels present the corresponding profiles for US stars. Filled blue crosses with dotted blue lines represent metal-rich stars and their best-fit models, while green crosses with solid green lines represent metal-poor stars and their fits. The metallicity subpopulations of LS stars behave similarly, showing no significant differences in their velocity dispersion and anisotropy profiles. For US stars, the velocity dispersion profiles of metal-rich and metal-poor stars are also similar, though subtle differences result in the metal-rich population being marginally more isotropic than the metal-poor population.

\subsection{Angular momentum}\label{subsec:angularmomentum}

We investigated the dispersion of the angular momentum ($\sigma_{L_z}$) as a function of radius. This parameter, recently introduced by \cite{aros2025}, is defined as $L_z = v_T \times r$, where $v_t$ is the tangential velocity and $r$ is the projected distance from the cluster center.

As discussed in \cite{aros2025}, $\sigma_{L_z}$ is connected with velocity anisotropy, since differences in the radial and tangential components of the velocity dispersion become apparent in the angular momentum distribution. In particular, at equivalent distances, stars on radial orbits display diminished angular momentum compared to those on more circular paths. 

\begin{figure}
    \centering
    \includegraphics[width=0.75\linewidth]{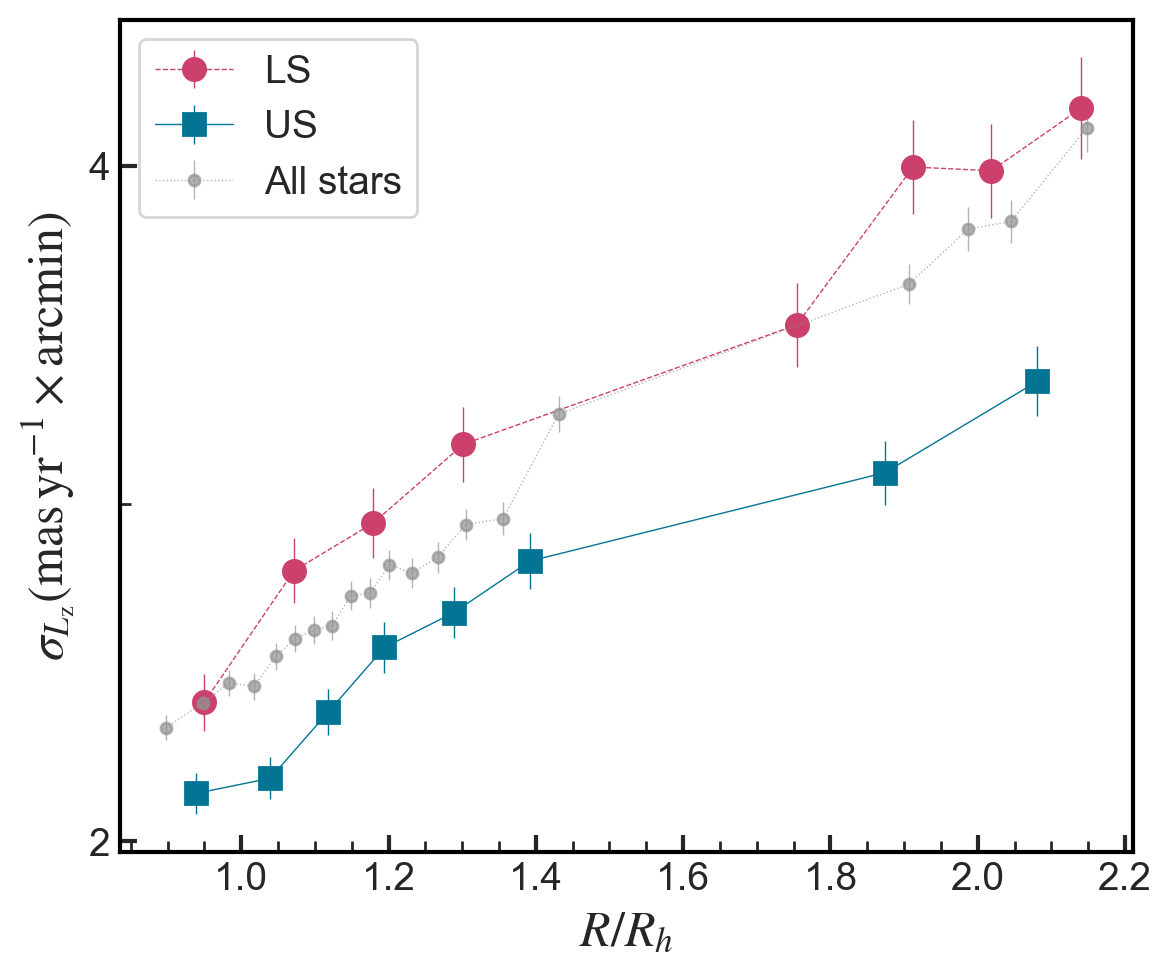}
    \caption{Radial profiles of the projected angular momentum dispersion $\sigma_{L_z}$ for all stars (gray), LS population (red), and US population (blue).}
    \label{fig:angularmomentum}
\end{figure}

Figure \ref{fig:angularmomentum} shows the dispersion of angular momentum as a function of radial distance for the multiple populations. The LS population consistently exhibits larger $\sigma_{L_z}$ values compared to the US population across the analyzed radial range, with these differences becoming more pronounced in the outermost regions. This behavior aligns with the increased radial anisotropy observed for US stars, as described in Section \ref{subsec:ani}. Radially anisotropic velocity distributions correspond to a reduced fraction of high-angular-momentum orbits, resulting in a narrower $L_z$ distribution for populations with greater anisotropy.

Since the proper motions used in this work are relative and do not retain the overall bulk rotation of the cluster (as the distortion correction process removes large-scale systematic motions), we performed a test to assess the impact of this effect on our results. We used Gaia absolute proper motions, which include the rotation signal, and measured the velocity dispersion, anisotropy, and angular momentum of stars in Omega Centauri over the same radial range considered in this study. We then removed the rotation component from these proper motions by fitting sinusoidal trends to the equatorial motion components as a function of position angle, following the method described in \cite{ziliotto2025a}, and repeated the measurements using the de-rotated motions. We found that the dispersion-related quantities — such as velocity dispersion, anisotropy, and angular momentum dispersion — were unaffected by the presence or absence of rotation.

\subsection{High-order moments of proper motions}

We further analyzed the distribution of radial and tangential proper motions by measuring the high-order moments h3 (skewness) and h4 (kurtosis). Skewness quantifies the asymmetry of the distribution, while kurtosis measures deviations from the Gaussian shape, with higher values indicating sharper peaks and more frequent extreme values.  

\begin{figure}
    \centering
    \includegraphics[width=1\linewidth]{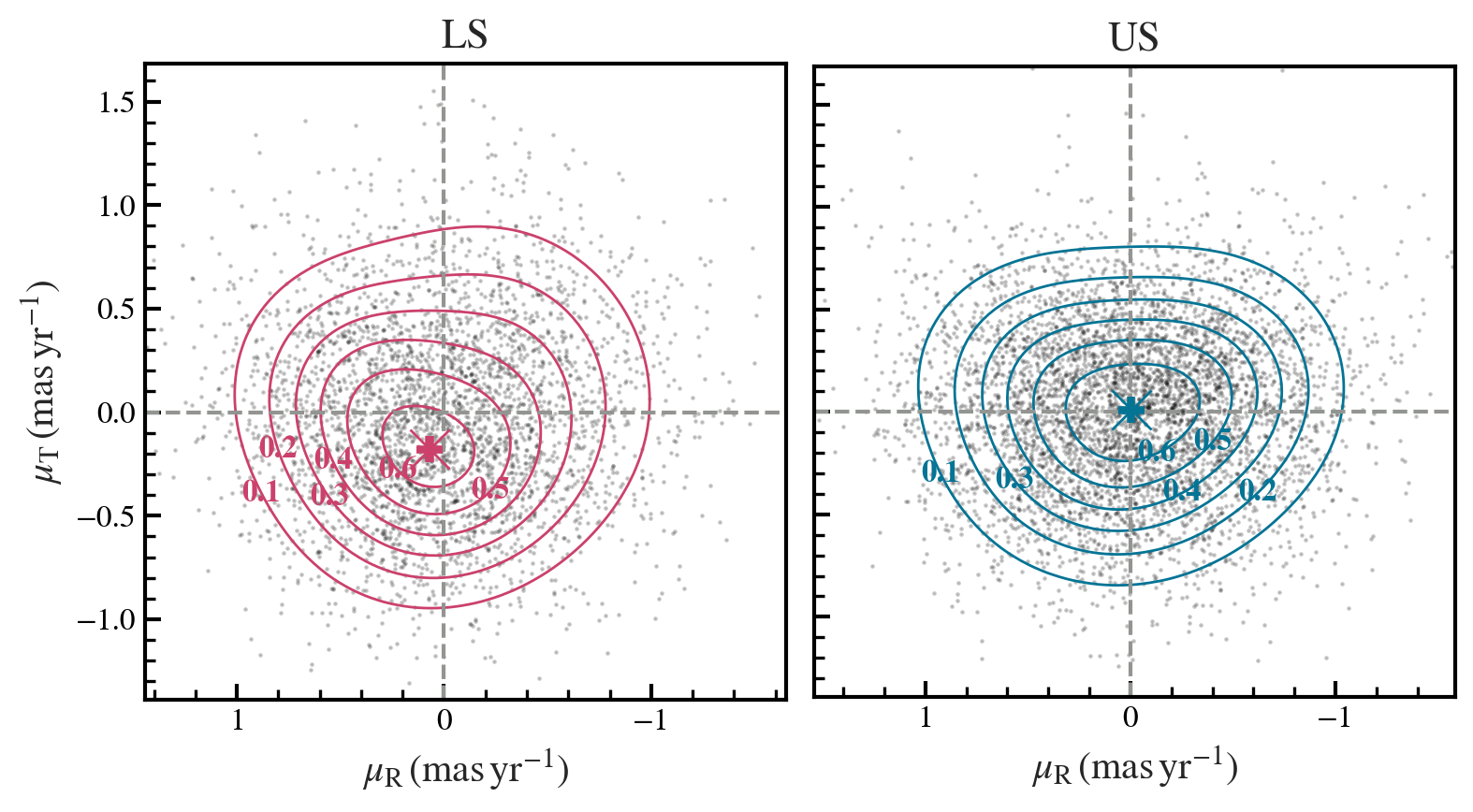}
    \caption{Tangential versus radial proper motions of LS (left panel) and US (right panel) stars. Contours indicate the probability density estimated by a Gaussian mixture model, highlighting the differences in statistical distributions and asymmetries between the two populations. The crosses indicate the centers of the distributions.}
    \label{fig:pms_LS_US}
\end{figure}

Skewness and kurtosis were measured using bootstrap resampling to account for single star velocity errors. For each radial bin, we performed 1000 bootstrap iterations where stars were resampled with replacement, and for each resampled star, a random value was drawn from a Gaussian distribution centered on the observed proper motion with a 1-$\sigma$ width equal to the measurement uncertainty. The skewness and kurtosis were computed for each bootstrap sample using \textit{scipy.stats} functions with bias correction, which adjusts the estimators to account for finite sample size effects, and the final values represent the mean across all bootstrap samples, with uncertainties given by the standard deviation of the bootstrap distribution.

Figure \ref{fig:pms_LS_US} shows the tangential and radial proper motion distributions for LS stars (left panel) and US stars (right panel). The distributions reveal notable differences between the two populations: LS stars exhibit more pronounced asymmetries in their proper motion patterns compared to the more symmetric distribution of US stars. The centers of the distributions are highlighted by the red (LS) and blue (US) crosses, which correspond to the peak of the probability grids. While US stars are consistent with being centered at around zero ($\mu_{\rm R}=-0.004 \pm 0.007, \, \mu_{\rm T}=+0.008 \pm 0.006$), LS stars are centered in $\mu_{\rm R} = 0.068 \pm 0.009$ and $\mu_{\rm T} = -0.180 \pm 0.008$, suggesting that LS stars have stronger systemic rotation with respect to US stars.

\begin{figure}
    \centering
    \includegraphics[width=0.75\linewidth]{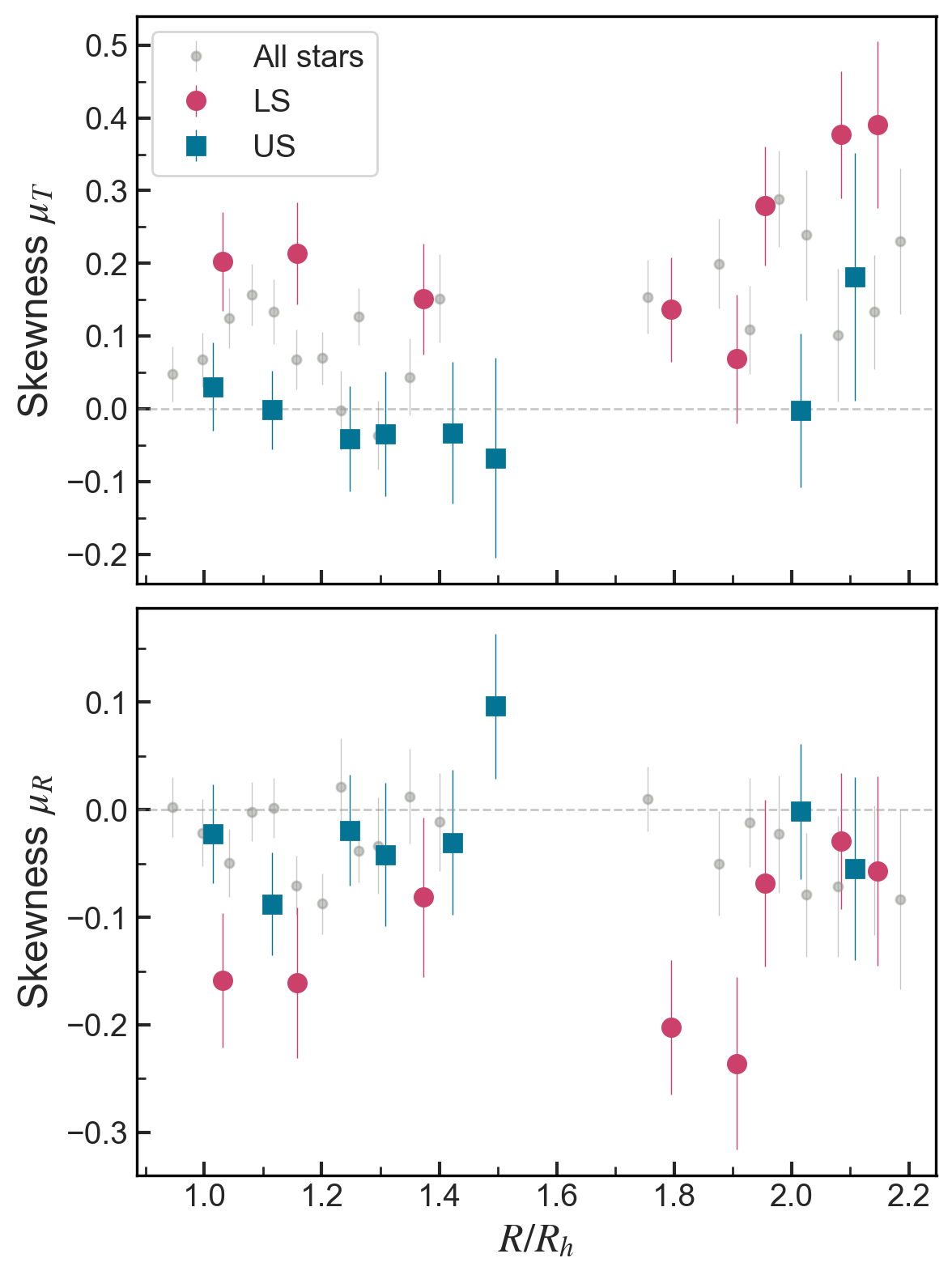}
    \caption{Radial variation of the skewness of the tangential (top) and radial (bottom) proper motion components. Gray points represent measurements for all stars, red circles for LS stars, and blue squares for US stars.}
    \label{fig:skewness}
\end{figure}

Figure \ref{fig:skewness} shows the radial variation of skewness in the tangential and radial proper motion components for LS (red points), US (blue points) and all stars. The US stars exhibit skewness values consistent with zero for both proper motion components. In contrast, LS stars show skewness with mean values of of $0.212 \pm 0.028$ for the tangential component and $-0.127 \pm 0.025$ for the radial component. This pattern is consistent with findings in the outer regions of $\omega$\,Centauri \citep{bellini2018a}, where MS-I stars (analogous to LS stars) showed skewness in their tangential proper motion distributions, while MS-II stars (analogous to US stars) exhibited no significant skewness.

The tangential kurtosis is nearly zero for both stellar populations. However, the radial component shows negative kurtosis for both populations, with LS stars exhibiting $-0.273 \pm 0.042$ and US stars showing $-0.295 \pm 0.034$, suggesting flatter-than-normal velocity distributions.

While the differences in skewness between LS and US stars provide additional evidence of the kinematic differences between these populations, its interpretation is not straightforward. When analyzing galaxy kinematics through line-of-sight velocity measurements, asymmetric velocity profiles are typically linked to internal rotation \citep[see, e.g.,][]{vandermarel1993,vandesande2017}. Nevertheless, when working with proper motion data, understanding the implications of skewness differences (or their absence) between tangential and radial velocity components becomes more challenging. This complexity is illustrated by \cite{ziliotto2025a}, who found that despite measurable differences in tangential velocity skewness between stellar generations in 47\,Tucanae, no corresponding significant rotational velocity differences are detected. Additional dynamical mechanisms may contribute to the observed velocity distribution characteristics. The interplay between intrinsic cluster dynamics and external Galactic tidal forces, combined with stellar escape processes, can modify the skewness properties of velocity distributions. To properly disentangle these effects, detailed N-body modeling will be required to trace how the initial kinematic properties of both populations evolve under various dynamical influences, ultimately determining their impact on the observed skewness patterns.

\subsection{Energy equipartition}

\begin{figure*}
    \centering
    \includegraphics[width=0.8\linewidth]{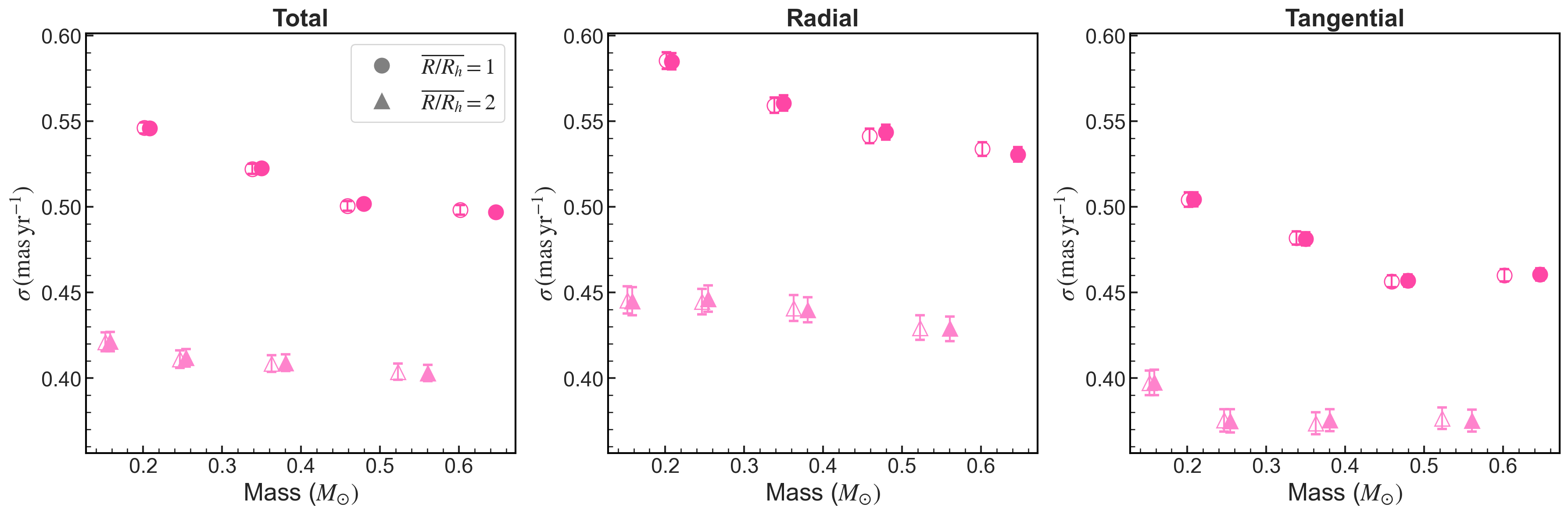}
    \caption{Velocity dispersion versus stellar mass for the total (left), radial (middle), and tangential (right) components. Circles and triangles correspond to the inner ($\bar{R}=1 R_h$) and outer ($\bar{R}=2 R_h$) regions, respectively. Filled symbols show the masses derived from the baseline isochrone, while open symbols show the masses obtained from the alternative model that includes the metallicity distribution from \cite{marino2011a} and a higher mean helium abundance (Y = 0.30).}
    \label{fig:disp_mass}
\end{figure*}

\begin{figure*}
    \centering
    \includegraphics[width=0.9\linewidth]{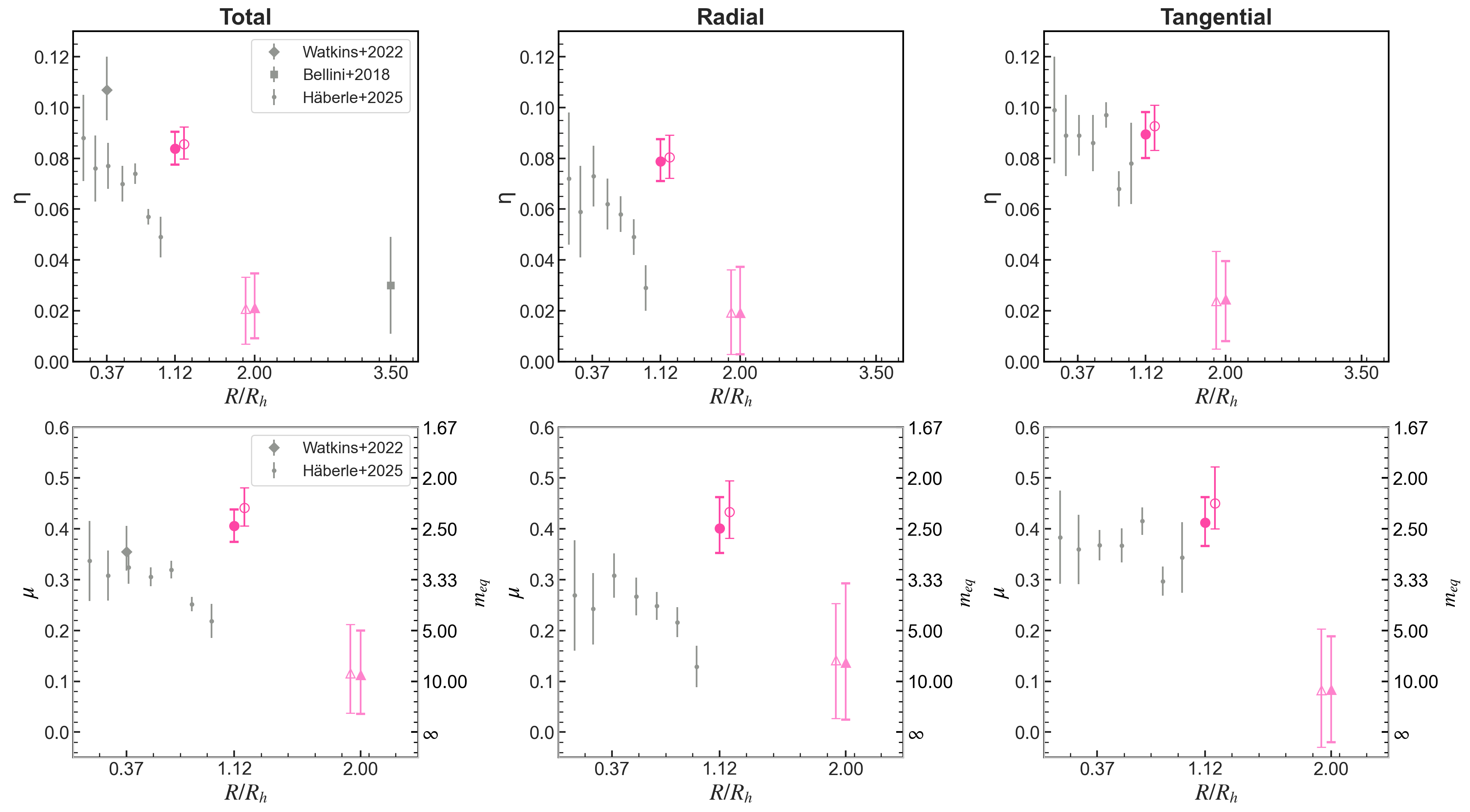}
    \caption{Best-fit parameters describing the degree of energy equipartition as a function of radial distance for the complete stellar sample. Top row: Equipartition parameter $\eta$. Bottom row: Mass parameters $\mu$ and $m_{eq}$. Left to right panels: Total, radial, and tangential velocity components. Points derived using the alternative, helium-enriched mass model (open markers) are slightly offset horizontally to improve the visibility of overlapping error bars.}
    \label{fig:equipartition}
\end{figure*}

We examined the evolution toward energy equipartition of the complete stellar sample in $\omega$\,Centauri within a mass range of $\sim0.15-0.7 M_{\odot}$. While we could distinguish multiple populations among lower and upper main-sequence stars, our sample is not continuous in mass, and thus differences in the degree of energy equipartition could not be measured with enough significance for US and LS stars.
We derived stellar masses by interpolating an isochrone with [$\alpha$/Fe]$=+0.4$, [Fe/H] = -1.55, Y = 0.25, and an age of 13.25 Gyr from the BaSTI database \citep{pietrinferni2021} to analyze the mass dependence of the velocity dispersion. Although $\omega$\,Centauri hosts a complex stellar population, a full treatment of the joint metallicity, helium, and light-element variations is not currently feasible due to the lack of detailed abundance distributions in this region and the limited coverage of existing stellar-evolution models. Following previous practice (e.g., \citealt{watkins2022}), we therefore adopted a representative single-metallicity, single-helium isochrone as our baseline model. To estimate the impact of population complexity on the inferred masses, we additionally computed a second set of masses using a more realistic approximation of the population mix, combining the metallicity distribution from \cite{marino2011a} with a higher mean helium abundance (Y = 0.30) to mimic the coexistence of helium-normal and helium-enhanced populations. The resulting masses (open symbols in Figure \ref{fig:disp_mass}) are slightly lower, but the derived equipartition parameters remain consistent with those from the baseline model (see Figure \ref{fig:equipartition}).

We measured the degree of energy equipartition by fitting two models, following the same procedure as \cite{ziliotto2025a}. The classical model assumes velocity dispersion scales with mass according to
\begin{equation}
    \sigma(m) = \sigma_{0}\left(\frac{m}{m_{0}}\right)^{-\eta},
    \label{eq:classic}
\end{equation}
where $\sigma_{0}$ is the velocity dispersion at the reference mass $m = m_0$, and $m_0$ is a scale mass, defined here as $m_0 = 1 M_{\odot}$. The parameter $\eta$ quantifies energy equipartition: $\eta=0$ indicates no equipartition, while $\eta=0.5$ represents full equipartition.

The second model incorporates an equipartition mass, $m_{\rm eq}$, originally defined by \cite{bianchini2016} as follows:
\begin{equation}
\sigma(m) =
    \begin{cases}
     \sigma_{0}\exp{(-0.5m/m_{\mathrm{eq}})}  & \mathrm{for \,} m \leq m_{\mathrm{eq}},\\
    \sigma_{\text{eq}}(m/m_{\mathrm{eq}})^{-0.5}  & \mathrm{for \,} m > m_{\mathrm{eq}},
    \end{cases}
    \label{eq
:bianchini}
\end{equation}
where the parameter $\sigma_\mathrm{eq}$ corresponds to the velocity dispersion at $m=m_{\rm eq}$ and is defined as $\sigma_{\mathrm{eq}} = \sigma_{0}\exp{(-0.5)}$. This formulation assumes complete equipartition for stars exceeding $m_{eq}$ and partial equipartition below this threshold. Lower $m_{eq}$ values indicate systems nearer to global equipartition. 

As in \cite{aros2023}, we defined $\mu = 1/m_{\mathrm{eq}}$.
This parametrization enables continuous parameter space exploration during fitting, including negative regimes. Systems with $\mu = 0$ exhibit no equipartition, while $\mu = 10$ approximates full equipartition. 

Figure \ref{fig:disp_mass} shows velocity dispersion versus stellar mass for the total, radial, and tangential components for all stars. We divided our sample into two radial bins, since the degree of energy equipartition can vary with radius, typically exhibiting higher values in the center of the cluster. The mean radii for each bin are labeled in the left panel. A trend consistent with energy equipartition is visible in all velocity components: lower-mass stars show smaller velocity dispersions than higher-mass stars.

The radial profiles of $\eta$, $\mu$, and $m_{\rm eq}$ are presented in Figure \ref{fig:equipartition}. 
Our derived values of $\eta$ are compared with those from \cite{watkins2022} and \cite{haberle2025a} for the central regions of $\omega$\,Centauri and from \cite{bellini2018a} at approximately $3.5 R_h$. Together, these measurements reveal a consistent radial decline in the degree of energy equipartition: \cite{watkins2022} find the highest values in the center, our results show intermediate values decreasing from $\eta = 0.07 \pm 0.01$ in the innermost bin to $\eta = 0.04 \pm 0.01$ in the outermost bin, and \cite{bellini2018a} report similarly low values at large radii. 
This declining trend is also evident in the radial and tangential components of the velocity dispersion (second and third columns of Figure \ref{fig:equipartition}).

\section{Summary and discussion}\label{sec:summary}

We have conducted a comprehensive investigation of the internal kinematics of multiple stellar populations in $\omega$\,Centauri using a combination of JWST NIRCam and HST observations. By analyzing stars in the previously unexplored intermediate radial region between $\sim$0.9 and 2.3 half-light radii from the cluster center, we characterized the kinematic properties of chemically distinct populations along the upper main sequence and among M dwarfs. Our photometric analysis identified multiple stellar populations using ChMs constructed from multiband {\it HST} and {\it JWST} photometry. We analyzed data from two main streams: LS stars, which exhibit light-element abundances similar to 1P GC stars, and US stars, which show the characteristic signatures of helium and nitrogen enhancement coupled with oxygen depletion typical of 2P populations. Both streams contain subpopulations with varying metallicities. This classification captures the complex subpopulation structure revealed in previous studies such as \cite{bellini2017a}, where the authors identified over a dozen distinct sequences in the core of $\omega$\,Centauri. 

Our kinematic analysis revealed significant differences in velocity anisotropy between the LS and US populations. The US stars display substantially stronger radial anisotropy compared to the LS stars, which remain nearly isotropic throughout the analyzed radial range. This anisotropy difference is primarily driven by variations in the tangential velocity dispersion component, with LS stars showing consistently higher tangential dispersions than US stars. 

\citet{cordoni2020a} investigated the internal kinematics of US and LS RGB stars within radial distances of $\sim$5.5 half-light radii by combining {\it HST} and {\it Gaia} proper motions and photometry. They found that US stars exhibit significantly stronger radial anisotropy than N-poor stars beyond $\sim$1.5 half-light radii. The results presented in this work support the conclusions of \citet{cordoni2020a} while being based on a substantially larger dataset covering radial distances between $\sim$1 and 2 half-light radii.
Notably, the kinematic behaviors of US and LS stars resemble those typically observed in dynamically young GCs, such as 47 Tucanae, where 2P stars exhibit stronger radial anisotropy than 1P stars \citep[][]{richer2013a, milone2018b, cordoni2020b, cordoni2025a, dalessandro2024, ziliotto2025a}. This similarity supports formation scenarios in which LS stars formed first, while the US stars originated from material polluted by more massive 1P stars \citep[e.g.,][]{marino2012a}.

When we further subdivided the US population based on their degree of chemical enrichment, we found a clear gradient in anisotropy behavior: The most chemically extreme US stars exhibit the highest radial anisotropy, intermediate streams fall between the extremes, and LS stars remain the most isotropic. Importantly,
 when comparing metal-rich and metal-poor stars within the same stream, we found moderate kinematic differences, suggesting that light-element abundance patterns rather than iron abundances are the primary drivers of the observed kinematic differences \citep[see also][]{cordoni2020a, haberle2025a}.

Our analysis of angular momentum properties revealed additional evidence of distinct dynamical behavior between the populations. The LS stars consistently exhibit higher angular momentum dispersion compared to US stars across the analyzed radial range. This pattern aligns with the increased radial anisotropy observed for US stars, as radially anisotropic velocity distributions correspond to a reduced fraction of high-angular-momentum orbits. The examination of higher-order moments of the proper motion distributions provides further evidence of kinematic differences: LS stars show significant skewness in both tangential and radial proper motion components, while US stars exhibit symmetric distributions consistent with zero skewness. Both populations display negative kurtosis in their radial components, indicating flatter-than-normal velocity distributions.

\citet{cordoni2020a} also found evidence of rotation among $\omega$Centauri stars, with the rotation amplitude decreasing from the center toward the outer regions \citep[see also][]{bellini2018a}. They concluded that N-rich stars exhibit different rotation patterns compared to N-poor stars, with a maximum tangential proper motion difference of $\sim$0.15mas~yr$^{-1}$ at radial distances of $\sim$1.5 half-light radii from the cluster center (see, e.g., their Figure 13). This conclusion is supported by our finding that the studied LS stars have higher average tangential proper motions than US stars, with $\Delta \mu_{\rm T} = 0.19 \pm 0.01$ mas yr$^{-1}$.

By analyzing all stars with available proper motions, we detected a low degree of energy equipartition ($\eta = 0.04-0.07$) across the analyzed mass range ($\sim0.15-0.7 M_{\odot}$), which decreases with increasing radial distance from the cluster center. This behavior is consistent with the general trend observed in GCs, where energy equipartition is most pronounced in the central regions and decreases toward the cluster periphery. 

\section*{Data availability}
Tables of the photometry, positions, and proper motions are only available in electronic form at the CDS via anonymous ftp to \url{cdsarc.u-strasbg.fr} (130.79.128.5) or via \url{http://cdsweb.u-strasbg.fr/cgi-bin/qcat?J/A+A/}.

\begin{acknowledgements}
      This research is based on observations made with the NASA/ESA \textit{Hubble Space Telescope} obtained from the Space Telescope Science Institute, which is operated by the Association of Universities for Research in Astronomy, Inc., under NASA contract NAS 5–26555. These observations are associated with programs GO-14118, GO-15857, GO-12580, GO-1677, GO-16247, GO-14759, and GO-9442.
      This work is based in part on observations made with the NASA/ESA/CSA James Webb Space Telescope. The data were obtained from the Mikulski Archive for Space Telescopes at the Space Telescope Science Institute, which is operated by the Association of Universities for Research in Astronomy, Inc., under NASA contract NAS 5-03127 for JWST. These observations are associated with program GO-6154. The authors acknowledge the team led by PI Anna de Graaff for developing their observing program with a zero-exclusive-access period.
      T. Z. acknowledges funding from the European Union’s Horizon 2020 research and innovation programme under the Marie Sklodowska-Curie Grant Agreement No. 101034319 and from the European Union – NextGenerationEU. This work has been funded by the European Union – NextGenerationEU RRF M4C2 1.1 (PRIN 2022 2022MMEB9W: `Understanding the formation of globular clusters with their multiple stellar generations', CUP C53D23001200006).
\end{acknowledgements}

\bibliographystyle{aa}
\bibliography{ms}
\end{document}